\documentclass[12pt]{article}%
\usepackage{ amsmath, graphics, rotating}%

\begin{document}
\begin{center} {\large\bf Positive Cosmological Constant and Quantum Theory}\end{center}

\vskip 1em \begin{center} {\large Felix M. Lev} \end{center}
\vskip 1em \begin{center} {\it Artwork Conversion Software Inc.,
1201 Morningside Drive, Manhattan Beach, CA 90266, USA
(Email:  felixlev314@gmail.com)} \end{center}

\begin{abstract} We argue that quantum theory should proceed not from
a spacetime background but from a Lie algebra, which is treated
as a symmetry algebra. Then the fact that the cosmological
constant is positive means not that the spacetime background is
curved  but that the de Sitter (dS) algebra as the symmetry
algebra is more relevant than the Poincare or anti de Sitter
ones. The physical interpretation of irreducible
representations (IRs) of the dS algebra is considerably
different from that for the other two algebras. One IR of the
dS algebra splits into independent IRs for a particle and its
antiparticle only when Poincare approximation works with a high
accuracy. Only in this case additive quantum numbers such as
electric, baryon and lepton charges are conserved, while at
early stages of the Universe they could not be conserved.
Another property of IRs of the dS algebra is that only fermions
can be elementary and there can be no neutral elementary
particles. The cosmological repulsion is a simple kinematical
consequence of dS symmetry on quantum level when quasiclassical
approximation is valid. Therefore the cosmological constant
problem does not exist and there is no need to involve dark
energy or other fields for explaining this phenomenon (in
agreement with a similar conclusion by Bianchi and Rovelli). 
\end{abstract}

\begin{flushleft} PACS: 02.10.De, 03.65.Ta, 11.30.Fs, 11.30.Ly\end{flushleft}

\begin{flushleft} Keywords: quantum theory; de Sitter invariance; cosmological constant problem \end{flushleft}

\begin{flushright} {\it In memory of Leonid Avksent'evich Kondratyuk}\end{flushright}

Our collaboration with Leonid Avksent'evich has considerably
enhanced my understanding of quantum theory. In particular, he
explained that the theory should not necessarily be based on a
local Lagrangian, and symmetry on quantum level means that
proper commutation relations are satisfied. I believe that the
present paper is in the spirit of these ideas.

\section{Introduction}
\label{S1}

The discovery of the cosmological repulsion (see e.g.,
\cite{Perlmutter,Melchiorri}) has ignited a vast discussion on how this
phenomenon should be interpreted. The majority of authors treat
this phenomenon as an indication that the cosmological constant
$\Lambda$ in General Relativity (GR) is positive and therefore
the spacetime background has a positive curvature. According to
References \cite{Spergel,Nakamura}, the observational data on the value
of $\Lambda$ define it with the accuracy better than 5\%.
Therefore the possibilities that $\Lambda=0$ or $\Lambda<0$ are
practically excluded. We argue in Sections
\ref{S2}--\ref{symmetry} that the notion of spacetime
background is not physical and, from the point of view of
quantum theory, one should proceed not from this notion but
from a symmetry algebra. Then the fact that in classical
approximation $\Lambda > 0$ is an indication that on quantum
level the de Sitter (dS) algebra is a more relevant symmetry
algebra than the Poincare or anti de Sitter (AdS) algebras. In
particular, elementary objects in quantum theory should be
described by irreducible representations (IRs) of the dS
algebra by Hermitian operators. In Sections \ref{S4}--\ref{S6}
we discuss mathematical properties of such IRs. Although there
exists a vast literature on IRs of the dS group and algebra,
their physical interpretation has not been widely discussed.
One of the main problems is that IRs of the dS algebra are
implemented on two Lorentz hyperboloids, not one as in the case
of the Poincare or AdS algebras. Therefore a problem arises how
IRs can be interpreted in terms of elementary particles and
antiparticles. In Reference \cite{lev1c} we have proposed an
interpretation that one IR of the dS algebra describes a
particle and its antiparticle simultaneously. In Section
\ref{S7} this analysis is considerably extended. In particular,
we show that  additive quantum numbers such as electric, baryon
and lepton charges can be only approximately conserved
quantities. It is also shown that only fermions can be
elementary and there can be no neutral elementary particles. In
Section \ref{S8} it is shown that cosmological repulsion is a
simple kinematical consequence of the dS symmetry on quantum
level when quasiclassical approximation is valid. For deriving
this result there is no need to involve spacetime background,
Riemannian geometry, de Sitter quantum field theory (QFT),
Lagrangians or other sophisticated methods. In other words, the
phenomenon of the cosmological repulsion can be naturally
explained on the basis of existing knowledge without involving
dark energy or other new fields (in agreement with a conclusion
by Bianchi and Rovelli in Reference \cite{Bianchi}). We tried
to make the presentation self-contained and make it possible
for readers to reproduce calculations without looking at other
papers.

\section{Remarks on the Cosmological Constant Problem}
\label{S2}

We would like to begin our presentation with a discussion of
the following well-known problem: How many independent
dimensionful constants are needed for a complete description of
nature? A paper \cite{Okun} represents a trialogue between
three well known scientists: M.J. Duff, L.B. Okun and G.
Veneziano. The results of their discussions are summarized as
follows: {\it LBO develops the traditional approach with three
constants, GV argues in favor of at most two (within
superstring theory), while MJD advocates zero.} According to
Reference \cite{W-units}, a possible definition of a
fundamental constant might be such that it cannot be calculated
in the existing theory. We would like to give arguments in
favor of the opinion of the first author in Reference
\cite{Okun}. One of our goals is to argue that the cosmological
constant cannot be a fundamental physical quantity.

Consider a measurement of a component of angular momentum. The
result depends on the system of units. As shown in quantum
theory, in units $\hbar/2=1$ the result is given by an integer
$0, \pm 1, \pm 2,...$. But we can reverse the order of units
and say that in units where the momentum is an integer $l$, its
value in $kg\cdot m^2/sec$ is $(1.05457162\cdot 10^{-34}\cdot
l/2)kg\cdot  m^2/sec$. Which of those two values has more
physical significance? In units where the angular momentum
components are integers, the commutation relations between the
components are
$$[M_x,M_y]=2iM_z\quad [M_z,M_x]=2iM_y\quad [M_y,M_z]=2iM_x$$
and they do not depend on any parameters. Then the meaning of
$l$ is clear: it shows how big the angular momentum is in
comparison with the minimum nonzero value 1. At the same time,
the measurement of the angular momentum in units $kg\cdot
m^2/sec$ reflects only a historic fact that at macroscopic
conditions on the Earth in the period between the 18th and 21st
centuries people measured the angular momentum in such units.

The fact that quantum theory can be written without the
quantity $\hbar$ at all is usually treated as a choice of units
where $\hbar=1/2$ (or $\hbar=1$). We believe that a better
interpretation of this fact is simply that quantum theory tells
us that physical results for measurements of the components of
angular momentum should be given in integers. Then the question
why $\hbar$ is as it is, is not a matter of fundamental physics
since the answer is: because we want to measure components of
angular momentum in $kg\cdot m^2/sec$.

Our next example is the measurement of velocity $v$. The fact
that any relativistic theory can be written without involving
$c$ is usually described as a choice of units where $c=1$. Then
the quantity $v$ can take only values in the range [0,1].
However, we can again reverse the order of units and say that
relativistic theory tells us that results for measurements of
velocity should be given by values in [0,1]. Then the question
why $c$ is as it is, is again not a matter of physics since the
answer is: Because we want to measure velocity in $m/sec$.

One might pose a question whether or not the values of $\hbar$
and $c$ may change with time. As far as $\hbar$ is concerned,
this is a question that if the angular momentum equals one then
its value in $kg\cdot  m^2/sec$ will always be $1.05457162\cdot
10^{-34}/2$ or not. It is obvious that this is not a problem of
fundamental physics but a problem how the units $(kg,m,sec)$
are defined. In other words, this is a problem of metrology and
cosmology. At the same time, the value of $c$ will always be
the same since the modern {\it definition} of meter is the
length which light passes during $(1/(3\cdot 10^8))sec$.

It is often believed that the most fundamental constants of
nature are $\hbar$, $c$ and the gravitational constant~$G$. The
units where $\hbar=c=G=1$ are called Planck units. Another well
known notion is the $c\hbar G$ cube of physical theories. The
meaning is that any relativistic theory should contain $c$, any
quantum theory should contain $\hbar$ and any gravitational
theory should contain $G$. However, the above remarks indicates
that the meaning should be the opposite. In particular,
relativistic theory {\it should not} contain $c$ and quantum
theory {\it should not contain} $\hbar$. The problem of
treating $G$ will be discussed below.

A standard phrase that relativistic theory becomes
non-relativistic one when $c\to\infty$ should be understood
such that if relativistic theory is rewritten in conventional
(but not physical!) units then $c$ will appear and one can take
the limit $c\to\infty$. A more physical description of the
transition is that all the velocities in question are much less
than unity. We will see in Section \ref{S8} that those
definitions are not equivalent. Analogously, a more physical
description of the transition from quantum to classical theory
should be that all angular momenta in question are very large
rather than $\hbar\to 0$.

Consider now what happens if we assume that de Sitter symmetry
is fundamental. For definiteness, we will discuss the dS
SO(1,4) symmetry and the same considerations can be applied to
the AdS symmetry SO(2,3). The dS space is a four-dimensional
manifold in the five-dimensional space defined by
\begin{equation}
x_1^2+x_2^2+x_3^2+x_4^2-x_0^2=R^2
\label{1}
\end{equation}
In the formal limit $R\to\infty$ the action of the dS group in
a vicinity of the point $(0,0,0,0,x_4= R)$ becomes the action
of the Poincare group on Minkowski space. In the literature,
instead of $R$, the cosmological constant $\Lambda=3/R^2$ is
often used. Then $\Lambda>0$ in the dS case and $\Lambda<0$ in
the AdS one. The dS space can be parameterized without using
the quantity $R$ at all if instead of $x_a$ ($a=0,1,2,3,4$) we
define dimensionless variables $\xi_a=x_a/R$. It is also clear
that  the elements of the SO(1,4) group do not depend on $R$
since they are products of conventional and hyperbolic
rotations. So the dimensionful value of $R$ appears only if one
wishes to measure coordinates on the dS space in terms of
coordinates of the flat five-dimensional space where the dS
space is embedded in. This requirement does not have a
fundamental physical meaning. Therefore the value of $R$
defines only a scale factor for measuring coordinates in the dS
space. By analogy with $c$ and $\hbar$, the question  why $R$
is as it is, is not a matter of fundamental physics since the
answer is: Because we want to measure distances in meters. In
particular, there is no guarantee that the cosmological
constant is really a constant, {\it i.e.}, does not change with
time. It is also obvious that if the dS symmetry is assumed
from the beginning then the value of $\Lambda$ has no relation
to the value of $G$.

If one assumes that spacetime background is fundamental then in
the spirit of GR it is natural to think that the empty
spacetime is flat, {\it i.e.}, that $\Lambda=0$ and this was
the subject of the well-known dispute between Einstein and de
Sitter. However, as mentioned in Section \ref{S1}, it is now
accepted that $\Lambda\neq 0$ and, although it is very small,
it is positive rather than negative. If we accept
parameterization of the dS space as in Equation (\ref{1}) then
the metric tensor on the dS space is
\begin{equation}
g_{\mu\nu}=\eta_{\mu\nu}-x_{\mu}x_{\nu}/(R^2+x_{\rho}x^{\rho})
\label{2}
\end{equation}
where $\mu,\nu,\rho = 0,1,2,3$, $\eta_{\mu\nu}$ is the diagonal
tensor with the components
$\eta_{00}=-\eta_{11}=-\eta_{22}=-\eta_{33}=1$ and a summation
over repeated indices is assumed. It is easy to calculate the
Christoffel symbols in the approximation where all the
components of the vector $x$ are much less than $R$:
$\Gamma_{\mu,\nu\rho}=-x_{\mu}\eta_{\nu\rho}/R^2$. Then a
direct calculation shows that in the non-relativistic
approximation the equation of motion for a single particle is
\begin{equation}
{\bf a}={\bf r}c^2/R^2
\label{3}
\end{equation}
where ${\bf a}$ and ${\bf r}$ are the acceleration and the
radius vector of the particle, respectively.

The fact that even a single particle in the Universe has a
nonzero acceleration might be treated as contradicting the law
of inertia but this law has been postulated only for Galilean
or Poincare symmetries and we have ${\bf a}=0$ in the limit
$R\to\infty$. A more serious problem is that, according to
standard experience, any particle moving with acceleration
necessarily emits gravitational waves, any charged particle
emits electromagnetic waves {\it etc.} Does this experience
work in the dS world? This problem is intensively discussed in
the literature (see e.g., References \cite{AkhmedovA,AkhmedovB,AkhmedovC} and
references therein). Suppose we accept that, according to GR,
the loss of energy in gravitational emission is proportional to
the gravitational constant. Then one might say that in the
given case it is not legitimate to apply GR since the constant
$G$ characterizes interaction between different particles and
cannot be used if only one particle exists in the world.
However, the majority of authors proceed from the assumption
that the empty dS space cannot be literally empty. If the
Einstein equations are written in the form $G_{\mu\nu}+\Lambda
g_{\mu\nu}=(8\pi G/c^4)T_{\mu\nu}$ where $T_{\mu\nu}$ is the
stress-energy tensor of matter then the case of empty space is
often treated as a vacuum state of a field with the
stress-energy tensor $T^{vac}_{\mu\nu}$ such that $(8\pi
G/c^4)T^{vac}_{\mu\nu}=-\Lambda g_{\mu\nu}$. This field is
often called dark energy. With such an approach one implicitly
returns to Einstein's point of view that a curved space cannot
be empty. Then the fact that $\Lambda\neq 0$ is treated as a
dark energy on the flat background. In other words, this is an
assumption that Poincare symmetry is fundamental while dS one
is emergent.

However, in this case a new problem arises. The corresponding
quantum theory is not renormalizable and with reasonable
cutoffs, the quantity $\Lambda$ in units $\hbar=c=1$ appears to
be of order $1/l_{P}^2=1/G$ where $l_P$ is the Planck length.
It is obvious that since in the above theory the only
dimensionful quantities in units $\hbar=c=1$ are $G$ and
$\Lambda$, and the theory does not have other parameters, the
result that $G\Lambda$ is of order unity seems to be natural.
However, this value of $\Lambda$ is at least by 120 orders of
magnitude greater than the experimental one. Numerous efforts
to solve this cosmological constant problem have not been
successful so far although many explanations have been
proposed.

Many physicists argue that in the spirit of GR, the theory
should not depend on the choice of the spacetime background (so
called a principle of background independence) and there should
not be a situation when the flat background is preferable (see
e.g., a discussion in Reference \cite{Rickles1} and references
therein). Moreover, although GR has been confirmed in several
experiments in Solar system, it is not clear whether it can be
extrapolated to cosmological distances. In other words, our
intuition based on GR with $\Lambda=0$ cannot be always correct
if $\Lambda\neq 0$. In Reference \cite{Bianchi} this point of
view is discussed in details. The authors argue that a general
case of Einstein's equation is when $\Lambda$ is present and
there is no reason to believe that a special case $\Lambda=0$
is preferable.

In summary, numerous attempts to resolve the cosmological
constant problem have not converged to any universally accepted
theory. All those attempts are based on the notion of spacetime
background and in the next section we discuss whether this
notion is physical.

\begin{sloppypar}
\section{Should Physical Theories Involve Spacetime Background?}
\label{S3}
\end{sloppypar}

From the point of view of quantum theory, any physical quantity
can be discussed only in conjunction with the operator defining
this quantity. For example, in standard quantum mechanics the
quantity $t$ is a parameter, which has the meaning of time only
in classical limit since there is no operator corresponding to
this quantity. The problem of how time should be defined on
quantum level is very difficult and is discussed in a vast
literature (see e.g., References \cite{Rosen,Rosen2,Rickles2} and
references therein). It has been also well known since the
1930s \cite{NW} that, when quantum mechanics is combined with
relativity, there is no operator satisfying all the properties
of the spatial position operator. In other words, the
coordinates cannot be exactly measured even in situations when
exact measurements are allowed by the non-relativistic
uncertainty principle. In the introductory section of the
well-known textbook \cite{BLP} simple arguments are given that
for a particle with mass $m$, the coordinates cannot be
measured with the accuracy better than the Compton wave length
${\hbar}/mc$. This fact is mentioned in practically every
textbook on quantum field theory (see e.g., Reference
\cite{Weinberg}). Hence, the exact measurement is possible only
either in the non-relativistic limit (when $c\to\infty$) or
classical limit (when ${\hbar}\to 0)$. There also exists a wide
literature where the meaning of space is discussed (see e.g.,
References \cite{Rickles1,Rosen,Rosen2,Rickles2}).

We accept a principle that any definition of a physical
quantity is a description how this quantity should be measured.
In quantum theory this principle has been already implemented
but we believe that it should be valid in classical theory as
well. From this point of view, one can discuss if {\it
coordinates of particles} can be measured with a sufficient
accuracy, while the notion of spacetime background, regardless
of whether it is flat or curved, does not have a physical
meaning. Indeed, this notion implies that spacetime coordinates
are meaningful even if they refer not to real particles but to
points of a manifold which exists only in our imagination.
However, such coordinates are not measurable. To avoid this
problem one might try to treat spacetime background as a
reference frame. Note that even in GR, which is a pure
classical ({\it i.e.}, non-quantum) theory, the meaning of
reference frame is not clear. In standard textbooks (see e.g.,
Reference \cite{LL}) the reference frame in GR is defined as a
collection of weightless bodies, each of which is characterized
by three numbers (coordinates) and is supplied by a clock. Such
a notion (which resembles ether) is not physical even on
classical level and for sure it is meaningless on quantum
level. There is no doubt that GR is a great achievement of
theoretical physics and has achieved great successes in
describing experimental data. At the same time, it is based on
the notions of spacetime background or reference frame, which
do not have a clear physical meaning. Therefore it is
unrealistic to expect that successful quantum theory of gravity
will be based on quantization of GR. The results of GR should
follow from quantum theory of gravity only in situations when
spacetime coordinates of {\it real bodies} is a good
approximation while in general the formulation of quantum
theory might not involve spacetime background at all.

One might take objection that coordinates of spacetime
background in GR can be treated only as parameters defining
possible gauge transformations while final physical results do
not depend on these coordinates. Analogously, although the
quantity $x$ in the Lagrangian density $L(x)$ is not
measurable, it is only an auxiliary tool for deriving equations
of motion in classical theory and constructing Hilbert spaces
and operators in quantum theory. After this construction has
been done, one can safely forget about background coordinates
and Lagrangian. In other words, a problem is whether
nonphysical quantities can be present at intermediate stages of
physical theories. This problem has a long history discussed in
a vast literature (see e.g., a discussion in References
\cite{Rickles1,Rosen,Rosen2,Rickles2}). Probably Newton was the first
who introduced the notion of spacetime background but, as noted
in a paper in Wikipedia, "Leibniz thought instead that space
was a collection of relations between objects, given by their
distance and direction from one another". We believe that at
the fundamental level unphysical notions should not be present
even at intermediate stages. So Lagrangian can be at best
treated as a hint for constructing a fundamental theory. As
stated in Reference \cite{BLP}, local quantum fields and
Lagrangians are rudimentary notion, which will disappear in the
ultimate quantum theory. Those ideas have much in common with
the Heisenberg S-matrix program and were rather popular till
the beginning of the 1970's. Although no one took objections
against those ideas, they are now almost forgotten in view of
successes of gauge theories.

In summary, although the most famous successes of theoretical
physics have been obtained in theories involving spacetime
background, this notion does not have a physical meaning.
Therefore a problem arises how to explain the fact that physics
seems to be local with a good approximation. In Section
\ref{S8} it is shown that the result given by Equation
(\ref{3}) is simply a consequence of dS symmetry on quantum
level when quasiclassical approximation works with a good
accuracy. For deriving this result there is no need to involve
dS space, metric, connection, dS QFT and other sophisticated
methods. The first step in our approach is discussed in the
next section.

\section{Symmetry on Quantum Level}
\label{symmetry}

If we accept that quantum theory should not proceed from
spacetime background, a problem arises how symmetry should be
defined on quantum level. In the spirit of Dirac's paper
\cite{Dir}, we postulate that on quantum level a symmetry means
that a system is described by a set of operators, which satisfy
certain commutation relations. We believe that for
understanding this Dirac's idea the following example might be
useful. If we define how the energy should be measured (e.g.,
the energy of bound states, kinetic energy {\it etc.}), we have
a full knowledge about the Hamiltonian of our system. In
particular, we know how the Hamiltonian should commute with
other operators. In standard theory the Hamiltonian is also
interpreted as an operator responsible for evolution in time,
which is considered as a classical macroscopic parameter. In
situations when this parameter is a good approximate parameter,
macroscopic transformations from the symmetry group
corresponding to the evolution in time have a meaning of
evolution transformations. However, there is no guarantee that
such an interpretation is always valid (e.g., at the very early
stage of the Universe). In general, according to principles of
quantum theory, self-adjoint operators in Hilbert spaces
represent observables but there is no requirement that
parameters defining a family of unitary transformations
generated by a self-adjoint operator are eigenvalues of another
self-adjoint operator. A well known example from standard
quantum mechanics is that if $P_x$ is the $x$ component of the
momentum operator then the family of unitary transformations
generated by $P_x$ is $exp(iP_xx/\hbar)$ where $x\in
(-\infty,\infty)$ and such parameters can be identified with
the spectrum of the position operator. At the same time, the
family of unitary transformations generated by the Hamiltonian
$H$ is $exp(-iHt/\hbar)$ where $t\in (-\infty,\infty)$ and
those parameters cannot be identified with a spectrum of a
self-adjoint operator on the Hilbert space of our system. In
the relativistic case the parameters $x$ can be formally
identified with the spectrum of the Newton-Wigner position
operator \cite{NW} but it is well known that this operator does
not have all the required properties for the position operator.
So, although the operators $exp(iP_xx/\hbar)$ and
$exp(-iHt/\hbar)$ are well defined, their physical
interpretation as translations in space and time is not always
valid (see also a discussion in Section \ref{S7}).

The {\it definition} of the dS symmetry on quantum level is
that the operators $M^{ab}$ ($a,b=0,1,2,3,4$, $M^{ab}=-M^{ba}$)
describing the system under consideration satisfy the
commutation relations {\it of the dS Lie algebra} so(1,4), {\it
i.e.},
\begin{equation}
[M^{ab},M^{cd}]=-i (\eta^{ac}M^{bd}+\eta^{bd}M^{ac}-
\eta^{ad}M^{bc}-\eta^{bc}M^{ad})
\label{4}
\end{equation}
where $\eta^{ab}$ is the diagonal metric tensor such that
$\eta^{00}=-\eta^{11}=-\eta^{22}=-\eta^{33}=-\eta^{44}=1$.
These relations do not depend on any free parameters. One might
say that this is a consequence of the choice of units where
$\hbar=c=1$. However, as noted above, any fundamental theory
should not involve the quantities $\hbar$ and $c$.

With such a definition of symmetry on quantum level, the dS
symmetry looks more natural than the Poincare symmetry. In the
dS case all the ten representation operators of the symmetry
algebra are angular momenta while in the Poincare case only six
of them are angular momenta and the remaining four operators
represent standard energy and momentum. If we define the
operators $P^{\mu}$ as $P^{\mu}=M^{4\mu}/R$ then in the formal
limit when $R\to\infty$, $M^{4\mu}\to\infty$ but the quantities
$P^{\mu}$ are finite, the relations (\ref{4}) will become the
commutation relations for representation operators of the
Poincare algebra such that the dimensionful operators $P^{\mu}$
are the four-momentum operators.

A theory based on the above definition of the dS symmetry on
quantum level cannot involve quantities which are dimensionful
in units $\hbar=c=1$. In particular, we inevitably come to
conclusion that the dS space, the gravitational constant and
the cosmological constant cannot be fundamental. The latter
appears only as a parameter replacing the dimensionless
operators $M^{4\mu}$ by the dimensionful operators $P^{\mu}$
which have the meaning of momentum operators only if $R$ is
rather large. Therefore the cosmological constant problem does
not arise at all but instead we have a problem why nowadays
Poincare symmetry is so good approximate symmetry. This is
rather a problem of cosmology but not quantum physics.

\section{IRs of the dS Algebra}
\label{S4}

If we accept dS symmetry on quantum level as described in the
preceding section, a question arises how elementary particles
in quantum theory should be defined. A discussion of numerous
controversial approaches can be found, for example in the
recent paper \cite{Rovelli}. Although particles are observables
and fields are not, in the spirit of QFT, fields are more
fundamental than particles, and a possible definition is as
follows \cite{Wein1}: {\it It is simply a particle whose field
appears in the Lagrangian. It does not matter if it's stable,
unstable, heavy, light---if its field appears in the Lagrangian
then it's elementary, otherwise it's composite.} Another
approach has been developed by Wigner in his investigations of
unitary irreducible representations (UIRs) of the Poincare
group \cite{Wigner}. In view of this approach, one might
postulate that a particle is elementary if the set of its wave
functions is the space of an IR of the symmetry group or Lie
algebra in the given theory. Since we do not accept approaches
based on spacetime background then by analogy with the Wigner
approach we accept that, {\it by definition}, elementary
particles in the dS invariant theory are described by IRs of
the dS algebra by Hermitian operators. For different reasons,
there exists a vast literature not on such IRs but on UIRs of
the dS group. References to this literature can be found e.g.,
in our papers \cite{lev1a,lev1b,lev1c,lev3} where we used the
results on UIRs of the dS group for constructing IRs of the dS
algebra by Hermitian operators. In this section we will
describe the construction proceeding from an excellent
description of UIRs of the dS group in a book by Mensky
\cite{Mensky}. The final result is given by explicit
expressions for the operators $M^{ab}$ in Equations (\ref{20})
and (\ref{21}). The readers who are not interested in technical
details can skip the derivation.

The elements of the SO(1,4) group will be described in the block form
\begin{equation}
g=\left\|\begin{array}{ccc}
g_0^0&{\bf a}^T&g_4^0\\
{\bf b}&r&{\bf c}\\
g_0^4&{\bf d}^T&g_4^4
\end{array}\right\|\
\label{5}
\end{equation}
where
\begin{equation}
\label{6}
{\bf a}=\left\|\begin{array}{c}a^1\\a^2\\a^3\end{array}\right\| \quad
{\bf b}^T=\left\|\begin{array}{ccc}b_1&b_2&b_3\end{array}\right\|
\quad r\in SO(3)
\end{equation}
and the subscript $^T$ means a transposed vector.

UIRs of the SO(1,4) group belonging to the principle series of
UIRs are induced from UIRs of the subgroup $H$ (sometimes
called ``little group'') defined as follows \cite{Mensky}. Each
element of $H$ can be uniquely represented as a product of
elements of the subgroups SO(3), $A$ and ${\bf T}$:
$h=r\tau_A{\bf a}_{\bf T}$ where
\begin{equation}
\tau_A=\left\|\begin{array}{ccc}
cosh(\tau)&0&sinh(\tau)\\
0&1&0\\
sinh(\tau)&0&cosh(\tau)
\end{array}\right\|\ \quad
{\bf a}_{\bf T}=\left\|\begin{array}{ccc}
1+{\bf a}^2/2&-{\bf a}^T&{\bf a}^2/2\\
-{\bf a}&1&-{\bf a}\\
-{\bf a}^2/2&{\bf a}^T&1-{\bf a}^2/2
\end{array}\right\|\
\label{7}
\end{equation}
The subgroup $A$ is one-dimensional and the three-dimensional
group ${\bf T}$ is the dS analog of the conventional
translation group (see e.g., Reference \cite{Mensky,Mielke}).
We believe it should not cause misunderstandings when 1 is used
in its usual meaning and when to denote the unit element of the
SO(3) group. It should also be clear when $r$ is a true element
of the SO(3) group or belongs to the SO(3) subgroup of the
SO(1,4) group. Note that standard UIRs of the Poincare group
are induced from the little group, which is a semidirect
product of SO(3) and four-dimensional translations and so the
analogy between UIRs of the Poincare and dS groups is clear.

Let $r\rightarrow \Delta(r;{\bf s})$ be an UIR of the group
SO(3) with the spin ${\bf s}$ and $\tau_A\rightarrow
exp(im_{dS}\tau)$ be a one-dimensional UIR of the group $A$,
where $m_{dS}$ is a real parameter. Then UIRs of the group $H$
used for inducing to the SO(1,4) group, have the form
\begin{equation}
\Delta(r\tau_A{\bf a}_{\bf T};m_{dS},{\bf s})=
exp(im_{dS}\tau)\Delta(r;{\bf s})
\label{8}
\end{equation}
We will see below that $m_{dS}$ has the meaning of the dS mass
and therefore UIRs of the SO(1,4) group are defined by the mass
and spin, by analogy with UIRs in Poincare invariant theory.

Let $G$=SO(1,4) and $X=G/H$ be the factor space (or coset
space) of $G$ over $H$. The notion of the factor space is well
known (see e.g., References \cite{Mackey,Naimark,Dixmier,Barut,Dobrev,Dobrev2,Mensky}). Each
element $x\in X$ is a class containing the elements $x_Gh$
where $h\in H$, and $x_G\in G$ is a representative of the class
$x$. The choice of representatives is not unique since if $x_G$
is a representative of the class $x\in G/H$ then $x_Gh_0$,
where $h_0$ is an arbitrary element from $H$, also is a
representative of the same class. It is well known that $X$ can
be treated as a left $G$ space. This means that if $x\in X$
then the action of the group $G$ on $X$ can be defined as
follows: if $g\in G$ then $gx$ is a class containing $gx_G$ (it
is easy to verify that such an action is correctly defined).
Suppose that the choice of representatives is somehow fixed.
Then $gx_G=(gx)_G(g,x)_H$ where $(g,x)_H$ is an element of $H$.
This element is called a factor.

The explicit form of the operators $M^{ab}$ depends on the
choice of representatives in the space $G/H$. As explained in
papers on UIRs of the SO(1,4) group (see e.g., Reference
\cite{Mensky}), to obtain the  possible closest analogy between
UIRs of the SO(1,4) and Poincare groups, one should proceed as
follows. Let ${\bf v}_L$ be a representative of the Lorentz
group in the factor space SO(1,3)/SO(3) (strictly speaking, we
should consider $SL(2,C)/SU(2)$). This space can be represented
as the well known velocity hyperboloid with the Lorentz
invariant measure
\begin{equation}
d\rho({\bf v})=d^3{\bf v}/v_0
\label{9}
\end{equation}
where $v_0=(1+{\bf v}^2)^{1/2}$. Let $I\in SO(1,4)$ be a matrix
which formally has the same form as the metric tensor $\eta$.
One can show (see e.g., Reference \cite{Mensky} for details)
that $X=G/H$ can be represented as a union of three spaces,
$X_+$, $X_-$ and $X_0$ such that $X_+$ contains classes ${\bf
v}_Lh$, $X_-$ contains classes ${\bf v}_LIh$ and $X_0$ has
measure zero relative to the spaces $X_+$ and $X_-$ (see also
Section \ref{S6}).

As a consequence, the space of UIR of the SO(1,4) group can be
implemented as follows. If $s$ is the spin of the particle
under consideration, then we use $||...||$ to denote the norm
in the space of UIR of the group SU(2) with the spin $s$. Then
the space of UIR is the space of functions $\{f_1({\bf
v}),f_2({\bf v})\}$ on two Lorentz hyperboloids with the range
in the space of UIR of the group SU(2) with the spin $s$ and
such that
\begin{equation}
\int\nolimits [||f_1({\bf v})||^2+
||f_2({\bf v})||^2]d\rho({\bf v}) <\infty
\label{10}
\end{equation}

It is well-known that positive energy UIRs of the Poincare and
AdS groups (associated with elementary particles) are
implemented on an analog of $X_+$ while negative energy UIRs
(associated with antiparticles) are implemented on an analog of
$X_-$. Since the Poincare and AdS groups do not contain
elements transforming these spaces to one another, the positive
and negative energy UIRs are fully independent. At the same
time, the dS group contains such elements (e.g., $I$
\cite{Mensky,Mielke}) and for this reason its UIRs can be
implemented only on the union of $X_+$ and $X_-$. Even this
fact is a strong indication that UIRs of the dS group cannot be
interpreted in the same way as UIRs of the Poincare and AdS
groups.

A general construction of the operators $M^{ab}$ is as follows.
We first define right invariant measures on $G=SO(1,4)$ and
$H$. It is well known (see e.g., References \cite{Mackey,Naimark,Dixmier,Barut}) that
for semisimple Lie groups (which is the case for the dS group),
the right invariant measure is simultaneously the left
invariant one. At the same time, the right invariant measure
$d_R(h)$ on $H$ is not the left invariant one, but has the
property $d_R(h_0h) = \Delta(h_0)d_R(h)$, where the number
function $h\rightarrow \Delta(h)$ on $H$ is called the module
of the group $H$. It is easy to show \cite{Mensky} that
\begin{equation}
\Delta(r\tau_A{\bf a}_{\bf T})=exp(-3\tau)
\label{11}
\end{equation}
Let $d\rho(x)$ be a measure on $X=G/H$ compatible with the
measures on $G$ and $H$. This implies that the measure on $G$
can be represented as $d\rho(x)d_R(h)$. Then one can show
\cite{Mensky} that if $X$ is a union of $X_+$ and $X_-$ then
the measure $d\rho(x)$ on each Lorentz hyperboloid coincides
with that given by Equation (\ref{9}). Let the representation
space be implemented as the space of functions $\varphi(x)$ on
$X$ with the range in the space of UIR of the SU(2) group such
that
\begin{equation}
\int\nolimits ||\varphi(x)||^2d\rho(x) <\infty
\label{12}
\end{equation}
Then the action of the representation operator $U(g)$
corresponding to $g\in G$ is defined as
\begin{eqnarray}
U(g)\varphi(x)=[\Delta((g^{-1},x)_H)]^{-1/2}
\Delta((g^{-1},x)_H;m_{dS},{\bf s})^{-1}\varphi(g^{-1}x)
\label{13}
\end{eqnarray}
One can directly verify that this expression defines a unitary
representation. Its irreducibility can be proved in several
ways (see e.g., Reference \cite{Mensky}).

As noted above, if $X$ is the union of $X_+$ and $X_-$, then
the representation space can be implemented as in Equation
(\ref{8}). Since we are interested in calculating only the
explicit form of the operators $M^{ab}$, it suffices to
consider only elements of $g\in G$ in an infinitely small
vicinity of the unit element of the dS group. In that case one
can calculate the action of representation operators on
functions having the carrier in $X_+$ and $X_-$ separately.
Namely, as follows from Equation (\ref{11}), for such $g\in G$,
one has to find the decompositions
\begin{equation}
g^{-1}{\bf v}_L={\bf v}'_Lr'(\tau')_A({\bf a}')_{\bf T}
\label{14}
\end{equation}
and
\begin{equation}
g^{-1}{\bf v}_LI={\bf v}"_LIr"(\tau")_A({\bf a}")_{\bf T}
\label{15}
\end{equation}
where $r',r"\in SO(3)$. In this expressions it suffices to
consider only elements of $H$ belonging to an infinitely small
vicinity of the unit element.

The problem of choosing representatives in the spaces
SO(1,3)/SO(3) or SL(2.C)/SU(2) is well known in standard
theory. The most usual choice is such that ${\bf v}_L$ as an
element of SL(2,C) is given by
\begin{equation}
{\bf v}_L=\frac{v_0+1+{\bf v}{\bf\sigma}}{\sqrt{2(1+v_0)}}
\label{16}
\end{equation}
Then by using a well known relation between elements of SL(2,C)
and SO(1,3) we obtain that ${\bf v}_L\in SO(1,4)$ is
represented by the matrix
\begin{equation}
{\bf v}_L=\left\|\begin{array}{ccc}
v_0&{\bf v}^T&0\\
{\bf v}&1+{\bf v}{\bf v}^T/(v_0+1)&0\\
0&0&1
\end{array}\right\|\
\label{17}
\end{equation}

As follows from Equations (\ref{8}) and (\ref{13}), there is no
need to know the expressions for $({\bf a}')_{\bf T}$ and
$({\bf a}")_{\bf T}$ in Equations (\ref{14}) and (\ref{15}). We
can use the fact \cite{Mensky} that if $e$ is the
five-dimensional vector with the components
$(e^0=1,0,0,0,e^4=-1)$ and $h=r\tau_A{\bf a}_{\bf T}$, then
$he=exp(-\tau)e$ regardless of the elements $r\in SO(3)$ and
${\bf a}_{\bf T}$. This makes it possible to easily calculate
$({\bf v}'_L,{\bf v}"_L,(\tau')_A,(\tau")_A)$ in Equations
(\ref{14}) and (\ref{15}). Then one can calculate $(r',r")$ in
these expressions by using the fact that the SO(3) parts of the
matrices $({\bf v}'_L)^{-1}g^{-1}{\bf v}_L$ and $({\bf
v}"_L)^{-1}g^{-1}{\bf v}_L$ are equal to $r'$ and $r"$,
respectively.

The relation between the operators $U(g)$ and $M^{ab}$ is as
follows. Let $L_{ab}$ be the basis elements of the Lie algebra
of the dS group. These are the matrices with the elements
\begin{equation}
(L_{ab})_d^c=\delta_d^c\eta_{bd}-\delta_b^c\eta_{ad}
\label{18}
\end{equation}
They satisfy the commutation relations
\begin{equation}
[L_{ab},L_{cd}]=\eta_{ac}L_{bd}-\eta_{bc}L_{ad}-
\eta_{ad}L_{bc}+\eta_{bd}L_{ac}
\label{19}
\end{equation}
Comparing Equations (\ref{4}) and (\ref{19}) it is easy to
conclude that the $M^{ab}$ should be the representation
operators of $-iL^{ab}$. Therefore if $g=1+\omega_{ab}L^{ab}$,
where a sum over repeated indices is assumed and the
$\omega_{ab}$ are such infinitely small parameters that
$\omega_{ab}=-\omega_{ba}$ then $U(g)=1+i\omega_{ab}M^{ab}$.

\begin{sloppypar}
We are now in position to write down the final expressions for
the operators $M^{ab}$. Their action on functions with the
carrier in $X_+$ has the form
\begin{eqnarray}
&&{\bf M}^{(+)}=l({\bf v})+{\bf s},\quad {\bf N}^{(+)}=-i v_0
\frac{\partial}{\partial {\bf v}}+\frac{{\bf s}\times {\bf v}}
{v_0+1}, \nonumber\\
&& {\bf B}^{(+)}=m_{dS} {\bf v}+i [\frac{\partial}{\partial {\bf v}}+
{\bf v}({\bf v}\frac{\partial}{\partial {\bf v}})+\frac{3}{2}{\bf v}]+
\frac{{\bf s}\times {\bf v}}{v_0+1},\nonumber\\
&& M_{04}^{(+)}=m_{dS} v_0+i v_0({\bf v}
\frac{\partial}{\partial {\bf v}}+\frac{3}{2})
\label{20}
\end{eqnarray}
where ${\bf M}=\{M^{23},M^{31},M^{12}\}$, ${\bf
N}=\{M^{01},M^{02},M^{03}\}$, ${\bf
B}=\{M^{41},M^{42},M^{43}\}$, ${\bf s}$ is the spin operator,
and ${\bf l}({\bf v})=-i{\bf v}\times \partial/\partial {\bf
v}$. At the same time, the action on functions with the carrier
in $X_-$ is given by
\begin{eqnarray}
&&{\bf M}^{(-)}=l({\bf v})+{\bf s},\quad {\bf N}^{(-)}=-i v_0
\frac{\partial}{\partial {\bf v}}+\frac{{\bf s}\times {\bf v}}
{v_0+1}, \nonumber\\
&& {\bf B}^{(-)}=-m_{dS} {\bf v}-i [\frac{\partial}{\partial {\bf v}}+
{\bf v}({\bf v}\frac{\partial}{\partial {\bf v}})+\frac{3}{2}{\bf v}]-
\frac{{\bf s}\times {\bf v}}{v_0+1},\nonumber\\
&& M_{04}^{(-)}=-m_{dS} v_0-i v_0({\bf v}
\frac{\partial}{\partial {\bf v}}+\frac{3}{2})
\label{21}
\end{eqnarray}
\end{sloppypar}
Note that the expressions for the action of the Lorentz algebra
operators on $X_+$ and $X_-$ are the same and they coincide
with the corresponding expressions for IRs of the Poincare
algebra. At the same time, the expressions for the action of
the operators $M^{4\mu}$ on $X_+$ and $X_-$ differ by sign.

In deriving Equations (\ref{20}) and (\ref{21}) we used only
the commutation relations (\ref{4}), no approximations have
been made and the results are exact. In particular, the dS
space, the cosmological constant and the Riemannian geometry
have not been involved at all. Nevertheless, the expressions
for the representation operators is all we need to have the
maximum possible information in quantum theory. If one defines
$m=m_{dS}/R$ and the operators $P^{\mu}$ as in the preceding
section then in the formal limit $R\to\infty$ we indeed obtain
the expressions for the operators of the IRs of the Poincare
algebra such that the Lorentz algebra operators are the same,
$E=mv_0$ and ${\bf P}=m{\bf v}$ where $E$ is the standard
energy operator and ${\bf P}$ is the standard momentum
operator. If we assume for definiteness that $m_{dS}>0$ then we
obtain positive energy and negative energy IRs of the Poincare
algebra on $X_+$ and $X_-$, respectively. It is obvious that in
that case $m$ is the standard mass in Poincare invariant
theory.

It is well known that in Poincare invariant theory the operator
$W=E^2-{\bf P}^2$ is the Casimir operator, {\it i.e.}, it
commutes with all the representation operators. According to
the well known Schur lemma in representation theory, all
elements in the space of IR are eigenvectors of the Casimir
operators with the same eigenvalue. In particular, they are the
eigenvectors of the operator $W$ with the \linebreak eigenvalue
$m^2$. As follows from Equation (\ref{4}), in the dS case the
Casimir operator of the second order is
$I_2=-1/2\sum_{ab}M_{ab}M^{ab}$ and this operator might be
treated as a dS analog of the mass operator squared. An
explicit calculation shows that for the operators given by
Equations (\ref{20}) and (\ref{21}) the numerical value of
$I_2$ is $m_{dS}^2-s(s+1)+9/4$.

\section{Absence of Weyl Particles in dS Invariant Theory}
\label{S5}

According to Standard Model, only massless Weyl particles can
be fundamental elementary particles in Poincare invariant
theory. Therefore a problem arises whether there exist analogs
of Weyl particles in dS invariant theory. In the preceding
section we have shown that the dS and Poincare masses are
related as $m_{dS}/R=m$. The dS mass is dimensionless while the
Poincare mass has the dimension $length^{-1}$. Since the
Poincare symmetry is a special case of the dS one, this fact is
in agreement with the observation in Section \ref{S2} that
dimensionful quantities cannot be fundamental. Let $l_C(m)$ be
the Compton wave length for the particle with the mass $m$.
Then one might think that, in view of the relation
$m_{dS}=R/l_C(m)$, the dS mass shows how many Compton wave
lengths are contained in the interval $(0,R)$. However, such an
interpretation of the dS mass means that we wish to interpret a
fundamental quantity $m_{dS}$ in terms of our experience based
on Poincare invariant theory. As already noted, the value of
$m_{dS}$ does not depend on any quantities having the dimension
$length$ or $length^{-1}$ and it is the Poincare mass which
implicitly depends on $R$. Let us assume for estimations that
the value of $R$ is $10^{28}cm$. Then even the dS mass of the
electron is of order $10^{39}$ and this might be treated as an
indication that the electron is not a true elementary particle.
Moreover, the present upper level for the photon mass is
$10^{-18}ev$ which seems to be an extremely tiny quantity.
However, the corresponding dS mass is of order $10^{15}$ and so
even the mass which is treated as extremely small in Poincare
invariant theory might be very large in dS invariant~theory.

In Poincare invariant theory, Weyl particles are characterized
not only by the condition that their mass is zero but also by
the condition that they have a definite helicity. Several
authors investigated dS and AdS analogs of Weyl particles
proceeding from covariant equations on the dS and AdS spaces,
respectively. For example, the authors of Reference
\cite{Ikeda} show that Weyl particles arise only when the dS or
AdS symmetries are broken to the Lorentz symmetry. At the level
of IRs, the existence of analogs of Weyl particles is known in
the AdS case. In Reference \cite{massless} we investigated such
analogs by using the results of References \cite{Evans} for
standard IRs of the AdS algebra ({\it i.e.}, IRs over the field
of complex numbers) and the results of Reference \cite{Braden}
for IRs of the AdS algebra over a Galois field. In the standard
case the minimum value of the AdS energy for massless IRs with
positive energy is $E_{min}=1+s$. In contrast with the
situation in Poincare invariant theory, where massless
particles cannot be in the rest state, the massless particles
in the AdS theory do have rest states and the value of the $z$
projection of the spin in such states can be $-s,-s+1, ..., s$ as
usual. However, for any value of energy greater than $E_{min}$,
the spin state is characterized only by helicity, which can
take the values either $s$ or $-s$, {\it i.e.}, we have the
same result as in Poincare invariant theory. In contrast with
IRs of the Poincare and dS algebra, IRs describing particles in
AdS theory belong to the discrete series of IRs and the energy
spectrum in them is discrete: $E=E_{min}, E_{min}+1,
..., \infty$. Therefore, strictly speaking, rest states do not
have measure zero. Nevertheless, the probability that the
energy is exactly $E_{min}$ is extremely small and therefore
there exists a correspondence between Weyl particles in
Poincare and AdS theories.

In Poincare invariant theory, IRs describing Weyl particles can
be constructing by analogy with massive IRs but the little group
is now E(2)instead of SO(3) (see e.g.,
section 2.5 in the textbook \cite{Weinberg}). The matter is
that the representation operators of the SO(3) group transform
rest states into themselves but for massless particles there
are no rest states. However, there exists another way of
getting massless IRs: one can choose the variables for massive
IRs in such a way that the operators of massless IRs can be
directly obtained from the operators of massive IRs in the
limit $m\to 0$. This construction has been described by several
authors (see e.g., References \cite{Kondratyuk,Fuda,FudaB,Ann} and
references therein) and the main stages are as follows. First,
instead of the $(0,1,2,3)$ components of vectors, we work with
the so called light front components $(+,-,1,2)$ where
$v^{\pm}=(v^0\pm v^3)/\sqrt{2}$ and analogously for other
vectors. We choose $(v^+,{\bf v}_{\bot})$ as three independent
components of the 4-velocity vector, where ${\bf
v}_{\bot}=(v_x,v_y)$. In these variables the measure (\ref{9})
on the Lorentz hyperboloid becomes $d\rho(v^+,{\bf
v}_{\bot})=dv^+d{\bf v}_{\bot}/v^+$. Instead of Equation
(\ref{16}) we now choose representatives of the SL(2,C)/SU(2)
classes as
\begin{equation}
v_L=\frac{1}{(v_0+v_z)^{1/2}}\left\|\begin{array}{cc}
v_0+v_z&0\\
v_x+iv_y&1
\end{array}\right\|\
\label{22}
\end{equation}
and by using the relation between the groups SL(2,C) and SO(1,3) we obtain that the form of this representative in the Lorentz group is
\begin{equation}
v_L=\left\|\begin{array}{cccc}
\sqrt{2}v^+&0&0&0\\
\frac{{\bf v}_{\bot}^2}{\sqrt{2}v^+}&\frac{1}{\sqrt{2}v^+}&\frac{v_x}{v^+}&\frac{v_y}{v^+}\\
\sqrt{2}v_x&0&1&0\\
\sqrt{2}v_y&0&0&1
\end{array}\right\|\
\label{23}
\end{equation}
where the raws and columns are in the order $(+,-,x,y)$.

By using the scheme described in the preceding section, we can
now calculate the explicit form of the representation operators
of the Lorentz algebra. In this scheme the form of these
operators in the IRs of the Poincare and dS algebras is the
same and in the case of the dS algebra the action is the same
for states with the carrier in $X_+$ and $X_-$. The results of
calculations are:
\begin{eqnarray}
&&M^{+-}=iv^+\frac{\partial}{\partial v^+}\quad M^{+j}=iv^+\frac{\partial}{\partial v^j}\quad M^{12}=l_z({\bf v}_{\bot})+s_z\nonumber\\
&&M^{-j}=-i(v^j\frac{\partial}{\partial v^+}+v^-\frac{\partial}{\partial v^j})-\frac{\epsilon_{jl}}{v^+}(s^l+v^ls_z)
\label{24}
\end{eqnarray}
where a sum over $j,l=1,2$ is assumed and $\epsilon_{jl}$ has
the components $\epsilon_{12}=-\epsilon_{21}=1$,
$\epsilon_{11}=\epsilon_{22}=0$. In Poincare invariant theories
one can define the standard four-momentum $p=mv$ and choose
$(p^+,{\bf p}_{\bot})$ as independent variables. Then the
expressions in Equation (\ref{24}) can be rewritten as
\begin{eqnarray}
&&M^{+-}=ip^+\frac{\partial}{\partial p^+}\quad M^{+j}=ip^+\frac{\partial}{\partial p^j}\quad M^{12}=l_z({\bf p}_{\bot})+s_z\nonumber\\
&&M^{-j}=-i(p^j\frac{\partial}{\partial p^+}+p^-\frac{\partial}{\partial p^j})-\frac{\epsilon_{jl}}{p^+}(ms^l+p^ls_z)
\label{25}
\end{eqnarray}
In dS invariant theory we can work with the same variables if
$m$ is defined as $m_{dS}/R$.

As seen from Equations (\ref{25}), only the operators $M^{-j}$
contain a dependence on the operators $s_x$ and $s_y$ but this
dependence disappears in the limit $m\to 0$. In this limit the
operator $s_z$ can be replaced by its eigenvalue $\lambda$
which now has the meaning of helicity. In Poincare invariant
theory the four-momentum operators $P^{\mu}$ are simply the
operators of multiplication by $p^{\mu}$ and therefore massless
particles are characterized only by one constant---helicity.

In dS invariant theory one can calculate the action of the
operators $M^{4\mu}$ by analogy with the calculation in the
preceding section. The actions of these operators on states
with the carrier in $X_+$ and $X_-$ differ only by sign and the
result for the actions on states with the carrier in $X_+$ is
\begin{eqnarray}
\label{26}
&&M^{4-}=m_{dS}v^-+i[v^-(v^+\frac{\partial}{\partial v^+}+v^j\frac{\partial}{\partial v^j}+\frac{3}{2})-\frac{\partial}{\partial v^+}]+
\frac{1}{v^+}\epsilon_{jl}v^js^l\nonumber\\
&&M^{4j}=m_{dS}v^j+i[v^j(v^+\frac{\partial}{\partial v^+}+v^l\frac{\partial}{\partial v^l}+\frac{3}{2})+\frac{\partial}{\partial v^j}]
-\epsilon_{jl}s^l\nonumber\\
&&M^{4+}=m_{dS}v^++iv^+(v^+\frac{\partial}{\partial v^+}+v^j\frac{\partial}{\partial v^j}+\frac{3}{2})
\end{eqnarray}
If we define $m=m_{dS}/R$ and $p^{\mu}=mv^{\mu}$ then for the
operators $P^{\mu}$ we have
\begin{eqnarray}
\label{27}
&&P^-=p^-+\frac{ip^-}{mR}(p^+\frac{\partial}{\partial p^+}+p^j\frac{\partial}{\partial p^j}+\frac{3}{2})-
\frac{im}{R}\frac{\partial}{\partial p^+}+\frac{1}{Rp^+}\epsilon_{jl}p^js^l\nonumber\\
&&P^j=p^j+\frac{ip^j}{mR}(p^+\frac{\partial}{\partial p^+}+p^l\frac{\partial}{\partial p^l}+\frac{3}{2})+
\frac{im}{R}\frac{\partial}{\partial p^j}-\frac{1}{R}\epsilon_{jl}s^l\nonumber\\
&&P^+=p^++\frac{ip^+}{mR}(p^+\frac{\partial}{\partial p^+}+p^j\frac{\partial}{\partial p^j}+\frac{3}{2})
\end{eqnarray}
Then it is clear that in the formal limit $R\to\infty$ we
obtain the standard Poincare result. However, when $R$ is
finite, the dependence of the operators $P^{\mu}$ on $s_x$ and
$s_y$ does not disappear. Moreover, in this case we cannot take
the limit $m\to 0$. Therefore we conclude that in dS theory
there are no Weyl particles, at least in the case when
elementary particles are described by IRs of the principle
series. Mensky conjectured \cite{Mensky} that massless
particles in the dS invariant theory might correspond to IRs of
the discrete series with $-im_{dS}=1/2$ but this possibility
has not been investigated. In any case, in contrast with the
situation in Poincare invariant theory, the limit of massive
IRs when $m\to 0$ does not give Weyl particles and moreover,
this limit does not exist.

\section{Other Implementations of IRs}
\label{S6}

In this section we will briefly describe two more
implementations of IRs of the dS algebra. The first one is
based on the fact that since SO(1,4)=SO(4)$A{\bf T}$ and
$H$=SO(3)$A{\bf T}$ \cite{Mensky}, there also exists a choice
of representatives which is probably even more natural than
those described above. Namely, we can choose as representatives
the elements from the coset space SO(4)/SO(3). Since the
universal covering group for SO(4) is SU(2)$\times$SU(2) and
for SO(3) --- SU(2), we can choose as representatives the
elements of the first multiplier in the product
SU(2)$\times$SU(2). Elements of SU(2) can be represented by the
points $u=({\bf u},u_4)$ of the three-dimensional sphere $S^3$
in the four-dimensional space as $u_4+i{\bf \sigma}{\bf u}$
where ${\bf \sigma}$ are the Pauli matrices and $u_4=\pm
(1-{\bf u}^2)^{1/2}$ for the upper and lower hemispheres,
respectively. Then the calculation of the operators is similar
to that described above and the results are as follows. The
Hilbert space is now the space of functions $\varphi (u)$ on
$S^3$ with the range in the space of the IR of the su(2)
algebra with the spin $s$ and such that
\begin{equation}
\int\nolimits ||\varphi(u)||^2du <\infty
\label{28}
\end{equation}
where $du$ is the SO(4) invariant volume element on $S^3$. The
explicit calculation  shows  that in this case the operators
have the form
\begin{eqnarray}
&&{\bf M}=l({\bf u})+{\bf s}\quad {\bf B}=i u_4
\frac{\partial}{\partial {\bf u}}-{\bf s} \quad M_{04}=(m_{dS} +3i/2)u_4+i u_4{\bf u}
\frac{\partial}{\partial {\bf u}} \nonumber\\
&& {\bf N}=-i [\frac{\partial}{\partial {\bf u}}-
{\bf u}({\bf u}\frac{\partial}{\partial {\bf u}})]
+(m_{dS} +3i/2){\bf u}-{\bf u}\times {\bf s}+u_4{\bf s}
\label{29}
\end{eqnarray}
Since Equations (\ref{10}), (\ref{20}) and (\ref{21}) on one
hand and Equations (\ref{28}) and (\ref{29}) on  the other  are
the  different implementations of one and   the   same
representation, there exists a unitary operator transforming
functions $f(v)$ into $\varphi (u)$ and operators
(\ref{20},\ref{21}) into operators (\ref{29}). For example in
the spinless case the operators (\ref{20}) and (\ref{29}) are
related to each other by a unitary transformation
\begin{equation}
\varphi (u)=exp(-im_{dS}lnv_0)v_0^{3/2}f(v)
\label{30}
\end{equation}
where the relation between the points of the upper hemisphere
and $X_+$ is ${\bf u}={\bf v}/v_0$ and $u_4=(1-{\bf
u}^2)^{1/2}$. The relation between the points of the lower
hemisphere and $X_-$ is ${\bf u}=-{\bf v}/v_0$ and
$u_4=-(1-{\bf u}^2)^{1/2}$.

The equator of $S^3$ where $u_4=0$ corresponds to $X_0$ and has
measure zero with respect to the upper and lower hemispheres.
For this reason one might think that it is of no interest for
describing particles in dS theory. Nevertheless, an interesting
observation is that while none of the components of $u$ has the
magnitude greater than unity, the set $X_0$ in terms of
velocities is characterized by the condition that $|{\bf v}|$
is infinitely large and therefore the standard Poincare
momentum ${\bf p}=m{\bf v}$ is infinitely large too. This poses
a question whether ${\bf p}$ always has a physical meaning.
From mathematical point of view Equation~(\ref{29}) might seem
more convenient than Equations (\ref{20}) and (\ref{21}) since
$S^3$ is compact and there is no need to break it into the
upper and lower hemispheres. In addition, Equation (\ref{29})
is an explicit implementation of the idea that since in dS
invariant theory all the variables $(x^1,x^2,x^3,x^4)$ are on
equal footing and so(4) is the maximal compact kinematical
algebra, the operators ${\bf M}$ and ${\bf B}$ do not depend on
$m_{dS}$. However, those expressions are not convenient for
investigating Poincare approximation since the Lorentz boost
operators ${\bf N}$ depend on $m_{dS}$.

Finally, we describe an implementation of IRs based on the
explicit construction of the basis in the representation space.
This construction is based on the method of su(2)$\times$su(2)
shift operators, developed by Hughes \cite{Hug} for
constructing UIRs of the group SO(5). It will be convenient for
us to  deal with the set of operators $({\bf J}',{\bf
J}'',R_{ij})$ ($i,j=1,2$) instead  of $M^{ab}$. Here ${\bf J}'$
and ${\bf J}"$ are two independent su(2) algebras ({\it i.e.},
$[{\bf J}',{\bf J}'']=0$). In each of them one chooses as the
basis the operators $(J_+,J_-,J_3)$ such that $J_1=J_++J_-$,
$J_2=-\imath (J_+-J_-)$ and the commutation relations have the
form
\begin{equation}
[J_3,J_+]=2J_+,\quad [J_3,J_-]=-2J_-,\quad [J_+,J_-]=J_3
\label{31}
\end{equation}
The commutation relations of the operators ${\bf J}'$ and
${\bf J}"$  with $R_{ij}$ have the form
\begin{eqnarray}
&&[J_3',R_{1j}]=R_{1j},\quad [J_3',R_{2j}]=-R_{2j},\quad
[J_3'',R_{i1}]=R_{i1},\nonumber\\
&& [J_3'',R_{i2}]=-R_{i2},\quad
[J_+',R_{2j}]=R_{1j},\quad [J_+'',R_{i2}]=R_{i1},\nonumber\\
&&[J_-',R_{1j}]=R_{2j},\quad [J_-'',R_{i1}]=R_{i2},\quad
[J_+',R_{1j}]=\nonumber\\
&&[J_+'',R_{i1}]=[J_-',R_{2j}]=[J_-'',R_{i2}]=0,\nonumber\\
\label{32}
\end{eqnarray}
and the commutation relations of the operators $R_{ij}$
with each other have the form
\begin{eqnarray}
&&[R_{11},R_{12}]=2J_+',\quad
[R_{11},R_{21}]=2J_+'',\nonumber\\
&& [R_{11},R_{22}]=-(J_3'+J_3''),\quad
[R_{12},R_{21}]=J_3'-J_3''\nonumber\\
&& [R_{11},R_{22}]=-2J_-'',\quad [R_{21},R_{22}]=-2J_-'
\label{33}
\end{eqnarray}
The relation between the sets $({\bf J}',{\bf J}",R_{ij})$ and $M^{ab}$  is given by
\begin{eqnarray}
&&{\bf M}=({\bf J}'+{\bf J}'')/2 \quad {\bf B}=({\bf J}'-{\bf J}'')/2
\quad M_{01}=i(R_{11}-R_{22})/2, \nonumber\\
&& M_{02}=(R_{11}+R_{22})/2 \quad
M_{03}=-i(R_{12}+R_{21})/2\quad  M_{04}=(R_{12}-R_{21})/2
\label{34}
\end{eqnarray}
Then it is easy to see that Equation (\ref{4}) follows from
Equations (\ref{32}--\ref{34}) and {\it vice versa}.

Consider the space of maximal  $su(2)\times su(2)$  vectors,
{\it i.e.},  such vectors $x$ that $J_+'x=J_+''x=0$. Then from
Equations (\ref{32}) and (\ref{33}) it follows that the
operators
\begin{eqnarray}
&&A^{++}=R_{11}\quad  A^{+-}=R_{12}(J_3''+1)-
J_-''R_{11}\quad A^{-+}=R_{21}(J_3'+1)-J_-'R_{11}\nonumber\\
&&A^{--}=-R_{22}(J_3'+1)(J_3''+1)+J_-''R_{21}(J_3'+1)+\nonumber\\
&&J_-'R_{12}(J_3''+1)-J_-'J_-''R_{11}
\label{35}
\end{eqnarray}
act invariantly on this space. The notations are related to the
property  that if $x^{kl}$  ($k,l>0$) is the maximal
su(2)$\times$su(2) vector and simultaneously the eigenvector of
operators $J_3'$ and $J_3"$ with the eigenvalues $k$ and $l$,
respectively, then  $A^{++}x^{kl}$ is  the  eigenvector  of
the  same operators with the values $k+1$ and $l+1$,
$A^{+-}x^{kl}$ - the eigenvector  with the values $k+1$ and
$l-1$, $A^{-+}x^{kl}$ - the eigenvector with the values  $k-1$
and $l+1$ and $A^{--}x^{kl}$ - the eigenvector with the values
$k-1$ and $l-1$.

The basis in the representation space can be explicitly
constructed assuming that there exists a vector $e^0$ which is
the maximal su(2)$\times$su(2) vector such that
\begin{equation}
J_3'e_0=0\quad J_3''e_0=se_0\quad A^{--}e_0=A^{-+}e_0=0\quad I_2e^0 =[m_{dS}^2-s(s+1)+9/4] e^0
\label{36}
\end{equation}
Then, as shown in Reference \cite{lev3}, the full basis of the
representation space consists of vectors
\begin{equation}
e_{ij}^{nr}=(J_-')^i (J_-'')^j(A^{++})^n(A^{+-})^re^0
\label{37}
\end{equation}
where $n=0,1,2,..., r$ can take only the values $0,1,...,2s$
and for the given $n$ and $s$, $i$ can take the values
$0,1,...,n+r$ and $j$ can take the values $0,1,...,n+2s-r$.

These results show that IRs of the dS algebra can be
constructed purely algebraically without involving analytical
methods of the theory of UIRs of the dS group. As shown in
Reference \cite{lev3}, this implementation is convenient for
generalizing standard quantum theory to a quantum theory over a
Galois field.

\section{Physical Interpretation of IRs of the dS Algebra}
\label{S7}

In Section \ref{S4}--\ref{S6} we discussed mathematical
properties of IRs of the dS algebra. In particular it has been
noted that they are implemented on two Lorentz hyperboloids,
not one as IRs of the Poincare algebra. Therefore the number of
states in IRs of the dS algebra is twice as big as in IRs of
the Poincare algebra. A problem arises whether this is
compatible with a requirement that any dS invariant theory
should become a Poincare invariant one in the formal limit
$R\to\infty$. Although there exists a wide literature on IRs of
the dS group and algebra, their physical interpretation has not
been widely discussed. Probably one of the reasons is that
physicists working on dS QFT treat fields as more fundamental
objects than particles (although the latter are observables
while the former are not).

In his book \cite{Mensky} Mensky notes that, in contrast with
IRs of the Poincare and AdS groups, IRs of the dS group
characterized by $m_{dS}$ and $-m_{dS}$ are unitarily
equivalent and therefore the energy sign cannot be used for
distinguishing particles and antiparticles. He proposes an
interpretation where a particle and its antiparticle are
described by the same IRs but have different space-time
descriptions (defined by operators intertwining IRs with
representations induced from the Lorentz group). Mensky shows
that in the general case his two solutions still cannot be
interpreted as a particle and its antiparticle, respectively,
since they are nontrivial linear combinations of functions with
different energy signs. However, such an interpretation is
recovered in Poincare approximation.

In view of the above discussion, it is desirable to give an
interpretation of IRs which does not involve spacetime. In
Reference \cite{lev1c} we have proposed an interpretation such
that one IR describes a particle and its antiparticle
simultaneously. In this section this analysis is extended.

\subsection{Problems with Physical Interpretation of IRs}

Consider first the case when the quantity $m_{dS}$ is very
large. Then, as follows from Equations (\ref{20}) and
(\ref{21}), the action of the operators $M^{4\mu}$ on states
localized on $X_+$ or $X_-$ can be approximately written as
$\pm m_{dS}v^{\mu}$, respectively. Therefore a question arises
whether the standard Poincare energy $E$ can be defined as
$E=M_{04}/R$. Indeed, with such a definition, states localized
on $X_+$ will have a positive energy while states localized on
$X_-$ will have a negative energy. Then a question arises
whether this is compatible with the standard interpretation of
IRs, according to which the following requirements should be
satisfied:

{\it Standard-Interpretation Requirements:} Each element of the
full representation space represents a possible physical state
for the given elementary particle. The representation
describing a system of $N$ free elementary particles is the
tensor product of the corresponding single-particle
representations.

Recall that the operators of the tensor product are given by
sums of the corresponding single-particle operators. For
example, if $M_{04}^{(1)}$ is the operator $M_{04}$ for
particle 1 and $M_{04}^{(2)}$ is the operator $M_{04}$ for
particle 2 then the operator $M_{04}$ for the free system
$\{12\}$ is given by $M_{04}^{(12)}=M_{04}^{(1)}+M_{04}^{(2)}$.
Here it is assumed that the action of the operator
$M_{04}^{(j)}$ ($j=1,2$) in the two-particle space is defined
as follows. It acts according to Equation (\ref{20}) or
(\ref{21}) over its respective variables while over the
variables of the other particle it acts as the identity
operator.

One could try to satisfy the standard interpretation as follows.

A) Assume that in Poincare approximation the standard energy
should be defined as $E = \pm M_{04}/R$ where the plus sign
should be taken for the states with the carrier in $X_+$ and as
the minus sign---for the states with the carrier in $X_-$. Then
the energy will always be positive definite.

B) One might say that the choice of the energy sign is only a
matter of convention. Indeed, to measure the energy of a
particle with the mass $m$ one has to measure its momentum
${\bf p}$ and then the energy can be defined not only as
$(m^2+{\bf p}^2)^{1/2}$ but also as $-(m^2+{\bf p}^2)^{1/2}$.
In that case the standard energy in the Poincare approximation
could be defined as $E = M_{04}/R$ regardless of whether the
carrier of the given state is in $X_+$ or $X_-$.

It is easy to see that either of the above possibilities is
incompatible with Standard-Interpretation Requirements.
Consider, for example, a system of two free particles in the
case when $m_{dS}$ is very large. Then with a high accuracy the
operators $M_{04}/R$ and ${\bf B}/R$ can be chosen diagonal
simultaneously.

Let us first assume that the energy should be treated according
to B). Then a system of two free particles with the equal
masses can have the same quantum numbers as the vacuum (for
example, if the first particle has the energy $E_0=(m^2+{\bf
p}^2)^{1/2}$ and momentum ${\bf p}$ while the second one has
the energy $-E_0$ and the momentum $-{\bf p}$) what obviously
contradicts experiment. For this and other reasons it is well
known that in Poincare invariant theory the particles should
have the same energy sign. Analogously, if the single-particle
energy is treated according to A) then the result for the
two-body energy of a particle-antiparticle system will
contradict experiment.

We conclude that IRs of the dS algebra cannot be interpreted in
the standard way since such an interpretation is physically
meaningless even in Poincare approximation. The above
discussion indicates that the problem we have is similar to
that with the interpretation of the fact that the Dirac
equation has solutions with both, positive and negative
energies.

As already noted, in Poincare and AdS theories there exist
positive energy IRs implemented on the upper hyperboloid and
negative energy IRs implemented on the lower hyperboloid. In
the latter case Standard-Interpretation Requirements are not
satisfied for the reasons discussed above. However, we cannot
declare such IRs unphysical and throw them away. In QFT quantum
fields necessarily contain both types of IRs such that positive
energy IRs are associated with particles while negative energy
IRs are associated with antiparticles. Then the energy of
antiparticles can be made positive after proper second
quantization. In view of this observation, we will investigate
whether IRs of the dS algebra can be interpreted in such a way
that one IR describes a particle and its antiparticle
simultaneously such that states localized on $X_+$ are
associated with a particle while states localized on $X_-$ are
associated with its~antiparticle.

By using Equation (\ref{10}), one can directly verify that the
operators (\ref{20}) and (\ref{21}) are Hermitian if the scalar
product in the space of IR is defined as follows. Since the
functions $f_1({\bf v})$ and $f_2({\bf v})$ in
Equation~(\ref{10}) have the range in the space of IR of the
su(2) algebra with the spin $s$, we can replace them by the
sets of functions $f_1({\bf v},j)$ and $f_2({\bf v},j)$,
respectively, where $j=-s,-s+1,...,s$. Moreover, we can combine
these functions into one function $f({\bf v},j,\epsilon)$ where
the variable $\epsilon$ can take only two values, say +1 or -1,
for the components having the carrier in $X_+$ or $X_-$,
respectively. If now $\varphi({\bf v},j,\epsilon)$ and
$\psi({\bf v},j,\epsilon)$ are two elements of our Hilbert
space, their scalar product is defined as
\begin{equation}
(\varphi,\psi)=\sum_{j,\epsilon}\int\nolimits
\varphi({\bf v},j,\epsilon)^*\psi({\bf v},j,\epsilon)
d\rho({\bf v}
\label{38})
\end{equation}
where the subscript $^*$ applied to scalar functions
means the usual complex conjugation.

At the same time, we use $^*$ to denote the operator adjoint to
a given one. Namely, if $A$ is the operator in our Hilbert
space then $A^*$ means the operator such that
\begin{equation}
(\varphi,A\psi)=(A^*\varphi,\psi)
\label{39}
\end{equation}
for all such elements $\varphi$ and $\psi$ that the left hand
side of this expression is defined.

Even in the case of the operators (\ref{20}) and (\ref{21}) we
can formally treat them as integral operators with some
kernels. Namely, if $A\varphi=\psi$, we can treat this relation
as
\begin{equation}
\sum_{j',\epsilon'}\int\nolimits
A({\bf v},j,\epsilon;{\bf v}',j',\epsilon')
\varphi({\bf v}',j',\epsilon')d\rho({\bf v}')=
\psi({\bf v},j,\epsilon)
\label{40}
\end{equation}
where in the general case the kernel $A({\bf v},j,\epsilon;{\bf
v}',j',\epsilon')$ of the operator $A$ is a distribution.

As follows from Equations (\ref{38}--\ref{40}), if $B=A^*$ then
the relation between the kernels of these operators is
\begin{equation}
B({\bf v},j,\epsilon;{\bf v}',j',\epsilon')=
A({\bf v}',j',\epsilon';{\bf v},j,\epsilon)^*
\label{41}
\end{equation}
In particular, if the operator $A$ is Hermitian then
\begin{equation}
A({\bf v},j,\epsilon;{\bf v}',j',\epsilon')^*=
A({\bf v}',j',\epsilon';{\bf v},j,\epsilon)
\label{42}
\end{equation}
and if, in addition, its kernel is real then the kernel is
symmetric, {\it i.e.},
\begin{equation}
A({\bf v},j,\epsilon;{\bf v}',j',\epsilon')=
A({\bf v}',j',\epsilon';{\bf v},j,\epsilon)
\label{43}
\end{equation}
In particular, this property is satisfied for the operators
$m_{dS} v_0$ and $m_{dS} {\bf v}$ in Equations (\ref{20}) and
(\ref{21}). At the same time, the operators
\begin{equation}
l({\bf v})\quad -i v_0\frac{\partial}{\partial {\bf v}}
\quad -i [\frac{\partial}{\partial {\bf v}}+{\bf v}({\bf v}\frac{\partial}{\partial
{\bf v}})+\frac{3}{2}{\bf v}]\quad -i v_0({\bf v}\frac{\partial}{\partial {\bf v}}+\frac{3}{2})
\label{44}
\end{equation}
which are present in Equations (\ref{20}) and (\ref{21}), are
Hermitian but have imaginary kernels. Therefore, as follows
from Equation (\ref{42}), their kernels are antisymmetric:
\begin{equation}
A({\bf v},j,\epsilon;{\bf v}',j',\epsilon')=-
A({\bf v}',j',\epsilon';{\bf v},j,\epsilon)
\label{45}
\end{equation}

In standard approach to quantum theory, the operators of
physical quantities act in the Fock space of the given system.
Suppose that the system consists of free particles and their
antiparticles. Strictly speaking, in our approach it is not
clear yet what should be treated as a particle or antiparticle.
The considered IRs of the dS algebra describe objects such that
$({\bf v}, j, \epsilon)$ is the full set of their quantum
numbers. Therefore we can define the annihilation and creation
operators $(a({\bf v},j,\epsilon),a({\bf v},j,\epsilon)^*)$ for
these objects. If the operators satisfy the anticommutation
relations then we require that
\begin{equation}
\{a({\bf v},j,\epsilon),a({\bf v}',j',\epsilon')^*\}=
\delta_{jj'}\delta_{\epsilon\epsilon'}v_0
\delta^{(3)}({\bf v}-{\bf v}')
\label{46}
\end{equation}
while in the case of commutation relations
\begin{equation}
[a({\bf v},j,\epsilon),a({\bf v}',j',\epsilon')^*]=
\delta_{jj'}\delta_{\epsilon\epsilon'}v_0
\delta^{(3)}({\bf v}-{\bf v}')
\label{47}
\end{equation}
In the first case, any two $a$-operators or any two $a^*$
operators anticommute with each other while in the second case
they commute with each other.

The problem of second quantization can now be formulated such
that IRs should be implemented as Fock spaces, {\it i.e.},
states and operators should be expressed in terms of the
$(a,a^*)$ operators. A possible implementation is as follows.
We define the vacuum state $\Phi_0$ such that is has a unit
norm and satisfies the requirement
\begin{equation}
a({\bf v},j,\epsilon)\Phi_0=0\quad \forall\,\, {\bf v},j,\epsilon
\label{vacuum}
\end{equation}
The image of the state $\varphi({\bf v},j,\epsilon)$ in the
Fock space is defined as
\begin{equation}
\varphi_F=\sum_{j,\epsilon}\int\nolimits \varphi({\bf v},j,\epsilon)a({\bf v},j,\epsilon)^*d\rho({\bf v})\Phi_0
\label{oneparticle}
\end{equation}
and the image of the operator with the kernel $A({\bf v},j,\epsilon;{\bf v}',j',\epsilon')$
in the Fock space is defined as
\begin{equation}
A_F=\sum_{j,\epsilon,j',\epsilon'}\int\nolimits\int\nolimits
A({\bf v},j,\epsilon;{\bf v}',j',\epsilon')
a({\bf v},j,\epsilon)^*a({\bf v}',j',\epsilon')
d\rho({\bf v})d\rho({\bf v}')
\label{48}
\end{equation}
One can directly verify that this is an implementation of IR in
the Fock space. In particular, the commutation relations in the
Fock space will be preserved regardless of whether the
$(a,a^*)$ operators satisfy commutation or anticommutation
relations and, if any two operators are adjoint in the
implementation of IR described above, they will be adjoint in
the Fock space as well. In other words, we have a $^*$
homomorphism of Lie algebras of operators acting in the space
of IR and in the Fock space.

We now require that in Poincare approximation the energy should
be positive definite. Recall that the operators (\ref{20}) and
(\ref{21}) act on their respective subspaces or in other words,
they are diagonal in the quantum number $\epsilon$.

Suppose that $m_{dS} > 0$ and consider the quantized operator
corresponding to the dS energy $M_{04}$ in Equation (\ref{20}).
In Poincare approximation, $M_{04}^{(+)}=m_{dS} v_0$ is fully
analogous to the standard free energy and therefore, as follows
from Equation (\ref{48}), its quantized form is
\begin{equation}
(M_{04}^{(+)})_F=m_{dS}\sum_{j}\int\nolimits v_0
a({\bf v},j,1)^*a({\bf v},j,1)d\rho({\bf v})
\label{49}
\end{equation}
This expression is fully analogous to the quantized Hamiltonian
in standard theory and it is well known that the operator
defined in such a way is positive definite.

Consider now the operator $M_{04}^{(-)}$. In Poincare
approximation its quantized form is
\begin{equation}
(M_{04}^{(-)})_F==-m_{dS}\sum_{j}\int\nolimits v_0
a({\bf v},j,-1)^*a({\bf v},j,-1)d\rho({\bf v})
\label{51}
\end{equation}
and this operator is negative definite, what is unacceptable.

One might say that the operators $a({\bf v},j,-1)$ and $a({\bf
v},j,-1)^*$ are ``nonphysical'': $a({\bf v},j,-1)$ is the
operator of object's annihilation with the negative energy, and
$a({\bf v},j,-1)^*$ is the operator of object's creation with
the negative energy.

We will interpret the operator $(M_{04}^{(-)})_F$ as that
related to antiparticles. In QFT the annihilation and creation
operators for antiparticles are present in quantized fields
with the coefficients describing negative energy solutions of
the corresponding covariant equation. This is an implicit
implementation of the idea that the creation or annihilation of
an antiparticle can be treated, respectively as the
annihilation or creation of the corresponding particle with the
negative energy. In our case this idea can be implemented
explicitly.

Instead of the operators $a({\bf v},j,-1)$ and $a({\bf
v},j,-1)^*$, we define new operators $b({\bf v},j)$ and $b({\bf
v},j)^*$. If $b({\bf v},j)$ is treated as the ``physical"
operator of antiparticle annihilation then, according to the
above idea, it should be proportional to $a({\bf v},-j,-1)^*$.
Analogously, if $b({\bf v},j)^*$ is the ``physical" operator of
antiparticle creation, it should be proportional to $a({\bf
v},-j,-1)$. Therefore
\begin{equation}
b({\bf v},j)=\eta (j)a({\bf v},-j,-1)^*\quad b({\bf v},j)^*= \eta (j)^*a({\bf v},-j,-1)
\label{53}
\end{equation}
where $\eta (j)$ is a phase factor such that
\begin{equation}
\eta (j)\eta(j)^*=1
\label{54}
\end{equation}
As follows from this relations, if a particle is characterized
by additive quantum numbers (e.g., electric, baryon or lepton
charges) then its antiparticle is characterized by the same
quantum numbers but with the minus sign. The transformation
described by Equations ({\ref{53}) and (\ref{54}) can also be
treated as a special case of the Bogolubov transformation
discussed in a wide literature on many-body theory (see, e.g.,
Chapter~10 in Reference \cite{Walecka} and references therein).

Since we treat $b({\bf v},j)$ as the annihilation operator and
$b({\bf v},j)^*$ as the creation one, instead of
Equation~(\ref{vacuum}) we should define a new vacuum state
${\tilde \Phi}_0$ such that
\begin{equation}
a({\bf v},j,1){\tilde \Phi}_0=b({\bf v},j){\tilde \Phi}_0=0\quad \forall\,\, {\bf v},j,
\label{55}
\end{equation}
and the images of states localized in $X_-$ should be defined as
\begin{equation}
\varphi_F^{(-)}=\sum_{j,\epsilon}\int\nolimits \varphi({\bf v},j,-1)b({\bf v},j)^*d\rho({\bf v})
{\tilde \Phi}_0
\label{newoneparticle}
\end{equation}
In that case the $(b,b^*)$ operators should be such that in the
case of anticommutation relations
\begin{equation}
\{b({\bf v},j),b({\bf v}',j')^*\}=
\delta_{jj'}v_0 \delta^{(3)}({\bf v}-{\bf v}'),
\label{56}
\end{equation}
and in the case of commutation relations
\begin{equation}
[b({\bf v},j),b({\bf v}',j')^*]=
\delta_{jj'}v_0 \delta^{(3)}({\bf v}-{\bf v}')
\label{57}
\end{equation}
We have to verify whether the new definition of the vacuum and
one-particle states is a correct implementation of IR in the
Fock space. A necessary condition is that the new operators
should satisfy the commutation relations of the dS algebra.
Since we replaced the $(a,a^*)$ operators by the $(b,b^*)$
operators only if $\epsilon=-1$, it is obvious from Equation
(\ref{48}) that the images of the operators (\ref{20}) in the
Fock space satisfy Equation (\ref{4}). Therefore we have to
verify that the images of the operators (\ref{21}) in the Fock
space also satisfy Equation (\ref{4}).

Consider first the case when the operators $a({\bf
v},j,\epsilon)$ satisfy the anticommutation relations. By using
Equation (\ref{53}) one can express the operators $a({\bf
v},j,-1)$ in terms of the operators $b({\bf v},j)$. Then it
follows from the condition (\ref{53}) that the operators
$b({\bf v},j)$ indeed satisfy Equation (\ref{56}). If the
operator $A_F$ is defined by Equation (\ref{48}) and is
expressed only in terms of the $(a,a^*)$ operators at
$\epsilon=-1$, then in terms of the $(b,b^*)$-operators it acts
on states localized in $X_-$ as
\begin{equation}
A_F=\sum_{j,j'}\int\nolimits\int\nolimits
A({\bf v},j,-1;{\bf v}',j',-1)\eta(j')\eta(j)^*b({\bf v},-j)b({\bf v}',-j')^*
d\rho({\bf v})d\rho({\bf v}')
\label{58}
\end{equation}
As follows from Equation (\ref{56}), this operator can be written as
\begin{equation}
A_F=C-\sum_{j,j'}\int\nolimits\int\nolimits
A({\bf v}',-j',-1;{\bf v},-j,-1)\eta(j)\eta(j')^*b({\bf v},j)^*b({\bf v}',j')
d\rho({\bf v})d\rho({\bf v}')
\label{59}
\end{equation}
where $C$ is the trace of the operator $A_F$:
\begin{equation}
C=\sum_j\int\nolimits A({\bf v},j,-1;{\bf v},j,-1)d\rho({\bf v})
\label{60}
\end{equation}
In general, $C$ is some indefinite constant. It can be
eliminated by requiring that all quantized operators should be
written in the normal form or by using another prescriptions.
The existence of infinities in the standard approach is the
well known problem and we will not discuss it. Therefore we
will always assume that if the operator $A_F$ is defined by
Equation (\ref{48}) then in the case of anticommutation
relations its action on states localized in $X_-$ can be
written as in Equation (\ref{59}) with $C=0$. Then, taking into
account the properties of the kernels discussed above, we
conclude that in terms of the $(b,b^*)$-operators the kernels
of the operators $(m_{dS}v)_F$ change their sign while the
kernels of the operators in Equation~(\ref{44}) remain the
same. In particular, the operator $(-m_{dS}v_0)_F$ acting on
states localized on $X_-$ has the same kernel as the operator
$(m_{ds}v_0)_F$ acting on states localized in $X_+$ has in
terms of the $a$-operators. This implies that in Poincare
approximation the energy of the states localized in $X_-$ is
positive definite, as well as the energy of the states
localized in $X_+$.

Consider now how the spin operator changes when the
$a$-operators are replaced by the $b$-operators. Since the spin
operator is diagonal in the variable ${\bf v}$, it follows from
Equation (\ref{59}) that the transformed spin operator will
have the same kernel if
\begin{equation}
s_i(j, j')=-\eta(j)\eta(j')^*s_i(-j',-j)
\label{61}
\end{equation}
where $s_i(j, j')$ is the kernel of the operator $s_i$. For the
$z$ component of the spin operator this relation is obvious
since $s_z$ is diagonal in $(j,j')$ and its kernel is
$s_z(j,j')=j\delta_{jj'}$.If we choose $\eta(j)=(-1)^{(s-j)}$
then the validity of Equation (\ref{61}) for $s=1/2$ can be
verified directly while in the general case it can be verified
by using properties of $3j$ symbols.

The above results for the case of anticommutation relations can
be summarized as follows. If we replace $m_{dS}$ by $-m_{dS}$
in Equation (\ref{21}) then the new set of operators
\begin{eqnarray}
&&{\bf M}'=l({\bf v})+{\bf s},\quad {\bf N}'=-i v_0
\frac{\partial}{\partial {\bf v}}+\frac{{\bf s}\times {\bf v}}
{v_0+1}, \nonumber\\
&& {\bf B}'=m_{dS} {\bf v}-i [\frac{\partial}{\partial {\bf v}}+
{\bf v}({\bf v}\frac{\partial}{\partial {\bf v}})+\frac{3}{2}{\bf v}]-
\frac{{\bf s}\times {\bf v}}{v_0+1},\nonumber\\
&& M_{04}'=m_{dS} v_0-i v_0({\bf v}
\frac{\partial}{\partial {\bf v}}+\frac{3}{2})
\label{62}
\end{eqnarray}
obviously satisfies the commutation relations (\ref{4}). The
kernels of these operators define quantized operators in terms
of the $(b,b^*)$-operators in the same way as the kernels of
the operators (\ref{20}) define quantized operators in terms of
the $(a,a^*)$-operators. In particular, in Poincare
approximation the energy operator acting on states localized in
$X_-$ can be defined as $E'=M_{04}'/R$ and in this
approximation it is positive definite.

At the same time, in the case of commutation relation the
replacement of the $(a,a^*)$-operators by the
$(b,b^*)$-operators is unacceptable for several reasons. First
of all, if the operators $a({\bf v},j,\epsilon)$ satisfy the
commutation relations (\ref{47}), the operators defined by
Equation (\ref{53}) will not satisfy Equation (\ref{57}). Also,
the r.h.s. of Equation (\ref{59}) will now have the opposite
sign. As a result, the transformed operator $M_{04}$ will
remain negative definite in Poincare approximation and the
operators (\ref{44}) will change their sign. In particular, the
angular momentum operators will no longer satisfy correct
commutation relations.

We have shown that if the definitions (\ref{vacuum}) and
(\ref{oneparticle}) are replaced by (\ref{55}) and
(\ref{newoneparticle}), respectively, then the images of both
sets of operators in Equation (\ref{20}) and Equation
(\ref{21}) satisfy the correct commutation relations in the
case of anticommutators. A question arises whether the new
implementation in the Fock space is equivalent to the IR
described in Section \ref{S4}. For understanding the essence of
the problem, the following very simple pedagogical example
might be useful.

Consider a representation of the SO(2) group in the space of
functions $f(\varphi)$ on the circumference $\varphi \in
[0,2\pi]$ where $\varphi$ is the polar angle and the points
$\varphi =0$ and $\varphi =2\pi$ are identified. The generator
of counterclockwise rotations  is $A=-id/d\varphi$ while the
generator of clockwise rotations is $id/d\varphi$. The equator
of the circumference contains two points, $\varphi=0$ and
$\varphi=\pi$ and has measure zero. Therefore we can represent
each $f(\varphi)$ as a superposition of functions with the
carriers in the upper and lower semi circumferences, $S_+$ and
$S_-$. The operators $A$ and $B$ are defined only on
differentiable functions. The Hilbert space $H$ contains not
only such functions but a set of differentiable functions is
dense in $H$. If a function $f(\varphi)$ is differentiable and
has the carrier in $S_+$ then $Af(\varphi)$ and $Bf(\varphi)$
also have the carrier in $S_+$ and analogously for functions
with the carrier in $S_-$. However, we cannot define a
representation of the SO(2) group such that its generator is
$A$ on functions with the carrier in $S_+$ and $B$ on functions
with the carrier in $S_-$ because a counterclockwise rotation
on $S_+$ should be counterclockwise on $S_-$ and analogously
for clockwise rotations. In other words, the actions of the
generator on functions with the carriers in $S_+$ and $S_-$
cannot be independent.

In the case of finite dimensional representations, any IR of a
Lie algebra by Hermitian operators can be always extended to an
UIR of the corresponding Lie group. In that case the UIR has a
property that any state is its cyclic vector {\it i.e.}, the
whole representation space can be obtained by acting by
representation operators on this vector and taking all possible
linear combinations. For infinite dimensional IRs this is not
always the case and there should exist conditions for IRs of
Lie algebras by Hermitian operators to be extended to
corresponding UIRs. This problem has been extensively discussed
in the mathematical literature (see e.g., References
\cite{Mackey,Naimark,Dixmier,Barut}). By analogy with finite dimensional IRs, one
might think that in the case of infinite dimensional IRs there
should exist an analog of the cyclic vector. In Section
\ref{S6} we have shown that for infinite dimensional IRs of the
dS algebra this idea can be explicitly implemented by choosing
a cyclic vector and acting on this vector by operators of the
enveloping algebra of the dS algebra. Therefore if IRs are
implemented as described in Section \ref{S4}, one might think
that the action of representation operators on states with the
carrier in $X_+$ should define its action on states with the
carrier in $X_-$.

\subsection{Example of Transformation Mixing Particles and Antiparticles}

We treated states localized in $X_+$ as particles and states
localized in $X_-$ as corresponding antiparticles. However, the
space of IR contains not only such states. There is no rule
prohibiting states with the carrier having a nonempty
intersection with both, $X_+$ and $X_-$. Suppose that there
exists a unitary transformation belonging to the UIR of the dS
group such that it transform a state with the carrier in $X_+$
to a state with the carrier in $X_-$. If the Fock space is
implemented according to Equations (\ref{vacuum}) and
(\ref{oneparticle}) then the transformed state will have the
form
\begin{equation}
\varphi_F^{(-)}=\sum_j\int\nolimits \varphi({\bf v},j)a({\bf v},j,-1)^*d\rho({\bf v})\Phi_0
\label{transform}
\end{equation}
while with the implementation in terms of the $(b,b^*)$
operators it should have the form (\ref{newoneparticle}). Since
the both states are obtained from the same state with the
carrier in $X_+$, they should be the same. However, they cannot
be the same. This is clear even from the fact that in Poincare
approximation the former has a negative energy while the latter
has a positive energy.

Our construction shows that the interpretation of states as
particles and antiparticles is not always consistent. We can
only guarantee that this interpretation is consistent when we
consider only states localized either in $X_+$ or in $X_-$ and
only transformations which do not mix such states. In quantum
theory there is a superselection rule (SSR) prohibiting states
which are superpositions of states with different electric,
baryon or lepton charges. In general, if states $\psi_1$ and
$\psi_2$ are such that there are no physical operators $A$ such
that $(\psi_2,A\psi_1)\neq 0$ then the SSR says that the state
$\psi=\psi_1+\psi_2$ is prohibited. The meaning of the SSR is
now widely discussed (see e.g., Reference \cite{Giulini} and
references therein). Since the SSR implies that the
superposition principle, which is a key principle of quantum
theory, is not universal, several authors argue that the SSR
should not be present in quantum theory. Other authors argue
that the SSR is only a dynamical principle since, as a result
of decoherence, the state $\psi$ will quickly disappear and so
it cannot be observable.

We now give an example of a transformation, which transform
states localized in $X_+$ to ones localized in $X_-$ and {\it
vice versa}. Let $I\in SO(1,4)$ be a matrix which formally
coincides with the metric tensor $\eta$. If this matrix is
treated as a transformation of the dS space, it transforms the
North pole $(0,0,0,0,x^4=R)$ to the South pole
$(0,0,0,0,x^4=-R)$ and {\it vice versa}. As already explained,
in our approach the dS space is not involved and in Sections
\ref{S4}--\ref{S6} the results for UIRs of the dS group have
been used only for constructing IRs of the dS algebra. This
means that the unitary operator $U(I)$ corresponding to $I$ is
well defined and we can consider its action without relating
$I$ to a transformation of the dS space.

If ${\bf v}_L$ is a representative defined by Equation
(\ref{17}) then it is easy to verify that $I{\bf v}_L=({-\bf
v})_L I$ and, as follows from Equation (\ref{13}), if $\psi_1$
is localized in $X_+$ then $\psi_2=U(I)\psi_1$ will be
localized in $X_-$. Therefore $U(I)$ transforms particles into
antiparticles and {\it vice versa}. In Section \ref{S3} we
argued that the notion of the spacetime background is
unphysical and that unitary transformations generated by
self-adjoint operators may not have a usual interpretation. The
example with $U(I)$ gives a good illustration of this point.
Indeed, if we work with the dS space, we might expect
that all unitary transformations corresponding to the elements
of the group SO(1,4) act in the space of IR only kinematically,
in particular they transform particles to particles and
antiparticles to antiparticles. However, in QFT in curved spacetime 
this is not the case. Nevertheless, this is not treated as an indication that 
standard notion of the dS space is not physical.
Although fields are not observable, in QFT in curved spacetime they 
are treated as fundamental and single-particle interpretations of field equations
are not tenable (moreover, some QFT theorists state that particles do not
exist). For example, as shown in References
\cite{Akhmedov2,Akhmedov2B,AkhmedovA,AkhmedovB,AkhmedovC}, solutions of fields equations are
superpositions of states which usually are interpreted as a
particle and its antiparticle, and in the dS space neither
coefficient in the superposition can be zero. This result is
compatible with the Mensky's one \cite{Mensky} described in the
beginning of this section. One might say that our result is in
agreement with those in dS QFT since UIRs of the dS group
describe not a particle or antiparticle but an object such that
a particle and its antiparticle are different states of this
object (at least in Poincare approximation). However, as noted above,
in dS QFT this is not treated as the fact that the dS space is
unphysical.

The matrix $I$ belongs to the component of unity of the group
SO(1,4). For example, the transformation $I$ can be obtained as
a product of rotations by 180 degrees in planes $(1,2)$ and
$(3,4)$. Therefore, $U(I)$ can be obtained as a result of
continuous transformations
$exp[i(M_{12}\varphi_1+M_{34}\varphi_2)]$ when the values of
$\varphi_1$ and $\varphi_2$ change from zero to $\pi$. Any
continuous transformation transforming a state with the carrier
in $X_+$ to the state with the carrier in $X_-$ is such that
the carrier should cross $X_0$ at some values of the
transformation parameters. As noted in the preceding section,
the set $X_0$ is characterized by the condition that the
standard Poincare momentum is infinite and therefore, from the
point of view of intuition based on Poincare invariant theory,
one might think that no transformation when the carrier crosses
$X_0$ is possible. However, as we have seen in the preceding
section, in variables $(u_1,u_2,u_3,u_4)$ the condition $u_4=0$
defines the equator of $S^3$ corresponding to $X_0$ and this
condition is not singular. So from the point of view of dS
theory, nothing special happens when the carrier crosses $X_0$.
We observe only either particles or antiparticles but not their
linear combinations because Poincare approximation works with a
very high accuracy and it is very difficult to perform
transformations mixing states localized in $X_+$ and $X_-$.

\subsection{Summary}

As follows from the above discussion, {\it objects belonging to
IRs of the dS algebra can be treated as particles or
antiparticles only if Poincare approximation works with a high
accuracy}. As a consequence, {\it the conservation of electric,
baryon and lepton charges can be only approximate}.

At the same time, our discussion shows that the approximation
when one IR of the dS algebra splits into independent IRs for a
particle and its antiparticle can be valid only in the case of
anticommutation relations. Since it is a reasonable requirement
that dS theory should become the Poincare one at certain
conditions, the above results show that {\it in dS invariant
theory only fermions can be elementary}.

Let us now consider whether there exist neutral particles in dS
invariant theory. In AdS and Poincare invariant theories,
neutral particles are described as follows. One first construct
a covariant field containing both IRs, with positive and
negative energies. Therefore the number of states is doubled in
comparison with the IR. However, to satisfy the requirement
that neutral particles should be described by real (not
complex) fields, one has to impose a relation between the
creation and annihilation operators for states with positive
and negative energies. Then the number of states describing a
neutral field again becomes equal to the number of states in
the IR. In contrast with those theories, IRs of the dS algebra
are implemented on both, upper and lower Lorentz hyperboloids
and therefore the number of states in IRs is twice as big as
for IRs of the Poincare and AdS algebras. Even this fact shows
that in dS invariant theory there can be no neutral particles
since it is not possible to reduce the number of states in IR.
Another argument is that, as follows from the above
construction, dS invariant theory is not $C$ invariant. Indeed,
$C$ invariance in standard theory means that representation
operators are invariant under the interchange of $a$-operators
and $b$-operators. However, in our case when $a$-operators are
replaced by $b$-operators, the operators (\ref{20}) become the
operators (\ref{62}). Those sets of operators coincide only in
Poincare approximation while in general the operators
$M^{4\mu}$ in Equations (\ref{20}) and (\ref{62}) are
different. Therefore a particle and its antiparticle are
described by different sets of operators. We conclude that {\it
in dS invariant theory neutral particles cannot be elementary}.

\section{dS Quantum Mechanics and Cosmological Repulsion}
\label{S8}

The results on IRs can be applied not only to elementary
particles but even to macroscopic bodies when it suffices to
consider their motion as a whole. This is the case when the
distances between the bodies are much greater that their sizes.
In this section we will consider the operators $M^{4\mu}$ not
only in Poincare approximation but taking into account dS
corrections. If those corrections are small, one can neglect
transformations mixing states on the upper and lower Lorentz
hyperboloids (see the discussion in the preceding section) and
describe the representation operators for a particle and its
antiparticle by Equations (\ref{20}) and (\ref{62}),
respectively.

We define $E=M_{04}/R$, ${\bf P}={\bf B}/R$ and $m=m_{dS}/R$.
Consider the non-relativistic approximation when $|{\bf v}|\ll
1$. If we wish to work with units where the dimension of
velocity is $m/sec$, we should replace ${\bf v}$ by ${\bf
v}/c$. If ${\bf p}=m{\bf v}$ then it is clear from the
expressions for ${\bf B}$ in Equations (\ref{20}) and
(\ref{62}) that ${\bf p}$ becomes the real momentum ${\bf P}$
only in the limit $R\to\infty$. Now by analogy with
nonrelativistic quantum mechanics, we {\it define} the position
operator ${\bf r}$ as $i\partial/\partial {\bf p}$. At this
stage we do not have any coordinate space yet. However, if we
assume that quasiclassical approximation is valid, we can treat
${\bf p}$ and ${\bf r}$ as usual vectors and neglect their
commutators. Then as follows from Equation (\ref{20})
\begin{equation}
{\bf P}= {\bf p}+mc{\bf r}/R\quad H = {\bf p}^2/2m +c{\bf p}{\bf r}/R
\label{64}
\end{equation}
where $H=E-mc^2$ is the classical non-relativistic Hamiltonian
and, as follows from Equations (\ref{62})
\begin{equation}
{\bf P}= {\bf p}-mc{\bf r}/R\quad H = {\bf p}^2/2m -c{\bf p}{\bf r}/R
\label{65}
\end{equation}
As follows from these expressions, in both cases
\begin{equation}
H({\bf P},{\bf r})=\frac{{\bf P}^2}{2m}-\frac{mc^2{\bf r}^2}{2R^2}
\label{66}
\end{equation}

The last term in Equation (\ref{66}) is the dS correction to
the non-relativistic Hamiltonian. It is interesting to note
that the non-relativistic Hamiltonian depends on $c$ although
it is usually believed that $c$ can be present only in
relativistic theory. This illustrates the fact mentioned in
Section \ref{S2} that the transition to non-relativistic theory
understood as $|{\bf v}|\ll 1$ is more physical than that
understood as $c\to\infty$. The presence of $c$ in Equation
(\ref{66}) is a consequence of the fact that this expression is
written in standard units. In non-relativistic theory $c$ is
usually treated as a very large quantity. Nevertheless, the
last term in Equation (\ref{66}) is not large since we assume
that $R$ is very large.

The result given by Equation (\ref{3}) is now a consequence of
the equations of motion for the Hamiltonian given by Equation
(\ref{66}). In our approach this result has been obtained
without using dS space and Riemannian geometry while the fact
that $\Lambda\neq 0$ should be treated not such that the
spacetime background has a curvature (since the notion of the
spacetime background is meaningless) but as an indication that
the symmetry algebra is the dS algebra rather than the Poincare
one. {\it Therefore for explaining the fact that $\Lambda\neq
0$ there is no need to involve dark energy or any other quantum
fields.}

Another way to show that our results are compatible with GR is
as follows. The well known result of GR is that if the metric
is stationary and differs slightly from the Minkowskian one
then in the non-relativistic approximation the curved spacetime
can be effectively described by a gravitational potential
$\varphi({\bf r})=(g_{00}({\bf r})-1)/2c^2$. We now express
$x_0$ in Equation (\ref{1}) in terms of a new variable $t$ as
$x_0=t+t^3/6R^2-t{\bf x}^2/2R^2$. Then the expression for the
interval becomes
\begin{equation}
ds^2=dt^2(1-{\bf r}^2/R^2)-d{\bf r}^2-
({\bf r}d{\bf r}/R)^2
\label{67}
\end{equation}
Therefore, the metric becomes stationary and $\varphi({\bf
r})=-{\bf r}^2/2R^2$ in agreement with Equation (\ref{66}).

Consider now a system of two free particles described by the
variables ${\bf p}_j$ and ${\bf r}_j$ ($j=1,2$). Define the
standard non-relativistic variables
\begin{eqnarray}
&&{\bf P}_{12}={\bf p}_1+{\bf p}_2
\quad {\bf q}_{12}=(m_2{\bf p}_1-m_1{\bf p}_2)/(m_1+m_2)\nonumber\\
&&{\bf R}_{12}=(m_1{\bf r}_1+m_2{\bf r}_2)/(m_1+m_2)\quad
{\bf r}_{12}={\bf r}_1-{\bf r}_2
\label{68}
\end{eqnarray}
Then if the particles are described by Equation (\ref{64}), the
two-particle operators ${\bf P}$ and ${\bf E}$ in the
non-relativistic approximation are given by
\begin{equation}
{\bf P}= {\bf P}_{12}+M{\bf R}_{12}/R,\quad
E = M+{\bf P}_{12}^2/2M +{\bf P}_{12}{\bf R}_{12}/R
\label{69}
\end{equation}
where
\begin{equation}
M = M({\bf q}_{12},{\bf r}_{12})=
m_1+m_2 +{\bf q}_{12}^2/2m_{12}+{\bf q}_{12}{\bf r}_{12}/R
\label{70}
\end{equation}
and $m_{12}$ is the reduced two-particle mass. Comparing
Equations (\ref{64}) and (\ref{70}), we conclude that $M$ has
the meaning of the two-body mass and therefore $M({\bf
q}_{12},{\bf r}_{12})$ is the internal two-body Hamiltonian. As
a consequence, in quasiclassical approximation the relative
acceleration is given by the same expression~(\ref{3}) but now
${\bf a}$ is the relative acceleration and ${\bf r}$ is the
relative radius vector.

The fact that two free particles have a relative acceleration
is well known for cosmologists who consider the dS symmetry on
classical level. This effect is called the dS antigravity. The
term antigravity in this context means that the particles
repulse rather than attract each other. In the case of the dS
antigravity the relative acceleration of two free particles is
proportional (not inversely proportional!) to the distance
between them. This classical result (which in our approach has
been obtained without involving dS space and Riemannian
geometry) is a special case of the dS symmetry on quantum level
when quasiclassical approximation works with a good accuracy.

For a system of two antiparticles the result is obviously the
same since Equation (\ref{65}) can be formally obtained from
Equation (\ref{64}) if $R$ is replaced by $-R$. At the same
time, in the case of a particle-antiparticle system a problem
with the separation of external and internal variables arises.
In any case the standard result can be obtained by using
Equation (\ref{66}).

Another problem discussed in the literature (see e.g.,
Reference \cite{Volovik} and references therein) is that
composite particles in the dS theory are unstable. As shown in
References \cite{lev1b,lev3}, if we assume that
non-relativistic approximation is valid but quasiclassical
approximation is not necessarily valid then the result
(\ref{70}) can be generalized as
\begin{equation}
H_{nr}=\frac{{\bf q}^2}{2m_{12}}+V_{dS},\quad V_{dS} = \frac{i}{R}({\bf q}\frac{\partial}{\partial
{\bf q}}+\frac{3}{2})
\label{71}
\end{equation}
where $H_{nr}$ is the non-relativistic internal Hamiltonian and
${\bf q}={\bf q}_{12}$. In spherical coordinates this
expression reads
\begin{equation}
H_{nr}=\frac{q^2}{2m_{12}} + \frac{i}{R}(q\frac{\partial}{\partial q}+\frac{3}{2})
\label{72}
\end{equation}
where $q=|{\bf q}|$. The operator (\ref{72}) acts in the space
of functions $\psi(q)$ such that
$\int_{0}^{\infty}|\psi(q)|^2q^2dq <\infty$ and the
eigenfunction $\psi_K$ of $E_{nr}$ with the eigenvalue $K$
satisfies the equation
\begin{equation}
q\frac{d\psi_K}{dq}=\frac{iRq^2}{m_{12}}\psi_K-(\frac{3}{2}+2iRK)\psi_K
\label{73}
\end{equation}
The solution of this equation is
\begin{equation}
\psi_K=\sqrt{\frac{R}{\pi}}q^{-3/2}exp(\frac{iRq^2}{2m_{12}}-2iRKlnq)
\label{74}
\end{equation}
and the normalization condition is
$(\psi_K,\psi_{K'})=\delta(K-K')$. The spectrum of the operator
$E_{nr}$ obviously belongs to the interval $(-\infty,\infty)$
and one might think that this is unacceptable. Suppose however
that $f(q)$ is a wave function of some state. As follows from
Equation (\ref{74}), the probability to have the value of the
energy $K$ in this state is defined by the coefficient $c(K)$
such that
\begin{equation}
c(K)=\sqrt{\frac{R}{\pi}}\int_{0}^{\infty}
exp(-\frac{iRq^2}{2m_{12}}+2iRKlnq)f(q)\sqrt{q}dq
\label{75}
\end{equation}
If $f(q)$ does not depend on $R$ and $R$ is very large then
$c(K)$ will practically be different from zero only if the
integrand in Equation (\ref{75}) has a stationary point $q_0$,
which is defined by the condition $K=q_0^2/2m_{12}$. Therefore,
for negative $K$, when the stationary point is absent, the
value of $c(K)$ will be exponentially small.

This result confirms that, as one might expect from Equation
(\ref{70}), the dS antigravity is not important for local
physics when $r\ll R$. At the same time, at cosmological
distances the dS antigravity is much stronger than any other
interaction (gravitational, electromagnetic {\it etc.}). Since
the spectrum of the energy operator is defined by its behavior
at large distances, this means that in the dS theory there are
no bound states. This does not mean that the theory is
unphysical since stationary bound states in standard theory
become quasistationary with a very large lifetime if $R$ is
large. For example, as shown in Equations (14) and (19) of
Reference \cite{Volovik}, a quasiclassical calculation of the
probability of the decay of the two-body composite system gives
that the probability equals $w=exp(-\pi\epsilon/H)$ where
$\epsilon$ is the binding energy and $H$ is the Hubble
constant. If we replace $H$ by $1/R$ and assume that
$R=10^{28}cm$ then for the probability of the decay of the
ground state of the hydrogen atom we get that $w$ is of order
$exp(-10^{35})$ {\it i.e.}, an extremely small value. This
result is in agreement with our remark after Equation
(\ref{75}).

In Reference \cite{lev1b} we discussed the following question.
In standard quantum mechanics the free Hamiltonian $H_0$ and
the full Hamiltonian $H$ are not always unitarily equivalent
since in the presence of bound states they have different
spectra. However, in the dS theory there are no bound states,
the free and full Hamiltonians have the same spectra and it is
possible to show that they are unitarily equivalent. Therefore
one can work in the formalism when interaction is introduced
not by adding an interaction operator to the free Hamiltonian
but by a unitary transformation of this operator. Such a
formalism might shed light on understanding of interactions in
quantum theory.

\section{Discussion and Conclusions}
\label{S9}

The experimental fact that $\Lambda >0$ might be an indication
that for some reasons nature prefers dS invariance vs. AdS
invariance ($\Lambda < 0$) and Poincare invariance
($\Lambda=0$). A question arises whether there exist
theoretical arguments explaining this fact. However, the
majority of authors treat $\Lambda >0$ as an anomaly since in
their opinion AdS invariance or Poincare invariance are more
preferable than dS invariance. One of the arguments is that dS
symmetry does not have a supersymmetric generalization in
contrast with the other two symmetries. Also, as argued by many
authors (see e.g., Reference \cite{Witten}), in QFT and its
generalizations (string theory, M-theory {\it etc.}) a theory
based on the dS algebra encounters serious difficulties. One of
the reasons is that IRs of the Poincare and AdS algebras
describing elementary particles are the lowest weight
representations where the Hamiltonian is positive definite. On
the other hand, as noted in the literature on IRs of the dS
algebra, the spectrum of any representation operator of this
algebra is symmetric relative to zero, {\it i.e.}, if $\lambda
> 0$ is an eigenvalue then $-\lambda$  also is an eigenvalue.
Polyakov \cite{Polyakov} believes that for this reason ``Nature
seems to abhor positive curvature and is trying to get rid of
it as fast as it can''.

Let us discuss this objection in greater details. IRs of the
Poincare and AdS algebras with the lowest weight are
implemented on the upper Lorentz hyperboloid in the velocity
space. Then the following question arises. If nature likes
Poincare and AdS symmetries then how should one treat the fact
that for any IR with the lowest weight $E_0>0$ there exists an
IR with the highest weight $-E_0$ on the lower hyperboloid?
Should one declare such IRs with negative energies unphysical
and throw them away? It is well known that the answer is ``no''
since local covariant objects needed for constructing QFT
(e.g., Dirac fields) necessarily contain IRs with lowest and
highest weight on equal footing. IRs with the lowest weight are
associated with particles while IRs with the highest weight
(and negative energies) are associated with antiparticles. Then
the problem of negative energies is solved by second
quantization after which both, the energies of particles and
antiparticles become positive. So, Polyakov's objection should
be understood such that only secondly quantized IRs with
positive energies are physical. If this is true then IRs of the
dS algebra are indeed unphysical.

In AdS and Poincare invariant theories, neutral particles are
described as follows. One first construct a covariant field
containing both IRs, with positive and negative energies.
Therefore the number of states is doubled in comparison with
the IR. However, to satisfy the requirement that neutral
particles should be described by real (not complex) fields, one
has to impose a relation between the creation and annihilation
operators for states with positive and negative energies. Then
the number of states describing a neutral field again becomes
equal to the number of states in the IR.

As shown in Reference \cite{Mensky} and other papers, IRs of
the dS algebra are implemented on both, upper and lower Lorentz
hyperboloids and therefore the number of states in IRs is twice
as big as for IRs of the Poincare and AdS algebras. A question
arises whether such a description is physical since the dS
theory should become the Poincare one when $R\to\infty$.
Another question is how one should distinguish particles and
antiparticles in the dS theory. In Reference \cite{lev1c} we
argued that the only possible physical interpretation of IRs is
such that they describe an object such that a particle and its
antiparticle are different states of this object in cases when
the wave function in the velocity space has a carrier on the
upper and lower Lorentz hyperboloids, respectively. In Section
\ref{S7} of the present paper we have shown that in dS
invariant theory
\begin{itemize}
\item Each state is either a particle or antiparticle only when one does not consider
transformations mixing states on the upper and lower
hyperboloids. Only in this case additive quantum numbers such
as electric, baryon and lepton charges are conserved. In
particular, they are conserved if Poincare approximation works
with a high accuracy.
\end{itemize}
In general, there is no superselection rule prohibiting states
which are superpositions of a particle and its antiparticle.
This shows that dS invariant theory implies a considerably new
understanding of the notion of particles and antiparticles. In
contrast with Poincare or AdS theories, for combining a
particle and its antiparticle together, there is no need to
construct a local covariant object since they are already
combined at the level of IRs.

We believe that this is an important argument in favor of dS
symmetry. Indeed, the fact that in AdS and Poincare invariant
theories a particle and its antiparticle are described by
different IRs means that they are different objects. Then a
problem arises why they have the same masses and spins but
opposite charges. In QFT this follows from the CPT theorem
which is a consequence of locality since {\it we construct}
local covariant fields from a particle and its antiparticle
with equal masses. A question arises what happens if locality
is only an approximation: In that case the equality of masses,
spins {\it etc.}, is exact or approximate? Consider a simple
model when electromagnetic and weak interactions are absent.
Then the fact that the proton and the neutron have the same
masses and spins has nothing to do with locality; it is only a
consequence of the fact that the proton and the neutron belong
to the same isotopic multiplet. In other words, they are simply
different states of the same object---the nucleon. We see, that
in dS invariant theories the situation is analogous. The fact
that a particle and its antiparticle have the same masses and
spins but opposite charges (in the approximation when the
notions of particles, antiparticles and charges are valid) has
nothing to do with locality or non-locality and is simply a
consequence of the fact that they are different states of the
same object since they belong to the same IR.

The non-conservation of the baryon and lepton quantum numbers
has been already considered in models of Grand Unification but
the electric charge has been always believed to be a strictly
conserved quantum number. In our approach all those quantum
numbers are not strictly conserved because in the case of de
Sitter symmetry transitions between a particle and its
antiparticle are not prohibited. The experimental data that
these quantum numbers are conserved reflect the fact that at
present Poincare approximation works with a very high accuracy.
As noted in Section \ref{S2}, the cosmological constant is not
a fundamental physical quantity and if the quantity $R$ is very
large now, there is no reason to think that it was large
always. This completely changes the status of the problem known
as ``baryon asymmetry of the Universe''.

Another consequence of our consideration is that {\it in dS
invariant theory only fermions can be elementary and there are
no neutral elementary particles}. The latter is obvious from
the fact that there is no way to reduce the number of states in
the IR. One might think that theories where the photon (and
also the graviton and the Higgs boson, if they exist) is not
elementary, cannot be physical. However, several authors
discussed models where the photon is composite; in particular,
in AdS theory it might be a composite state of Dirac's
singletons \cite{FF,CH,F}. An indirect confirmation of our
conclusions is that all known neutral particles are bosons.

One might say that a possibility that only fermions can be
elementary is not attractive since such a possibility would
imply that supersymmetry is not fundamental. There is no doubt
that supersymmetry is a beautiful idea. On the other hand, one
might say that there is no reason for nature to have both,
elementary fermions and elementary bosons since the latter can
be constructed from the former. A well know historical analogy
is that the simplest covariant equation is not the Klein-Gordon
equation for spinless fields but the Dirac and Weyl equations
for the spin 1/2 fields since the former is the equation of the
second order while the latter are the equations of the first
order.

We see that theories based on dS symmetry on one hand and on
AdS and Poincare symmetries on the other, are considerably
different. The problem with the interpretation of IRs of the dS
algebra has a clear analogy with the fact that the Dirac
equation has solutions with both, positive and negative
energies. As already noted, in modern quantum theory the latter
problem has been solved by second quantization. This is
possible because IRs of the Poincare and AdS algebras with
positive and negative energies are independent of each other.
On the contrary, one IR of the dS algebra contains states which
can be treated as particles and antiparticles only in some
approximations while in general, as we have seen in
Section~\ref{S7}, the second quantization formalism is not
sufficient for splitting all possible states into particle and
antiparticle ones.

Depending on their preferences, physicists may have different
opinions on this situation. As noted above, many physicists
treat the dS theory as unphysical and the fact that $\Lambda
>0$ as an anomaly. However, as noted in Section \ref{S1}, the
accuracy of experimental data is such that the possibilities
$\Lambda = 0$ or $\Lambda < 0$ are practically excluded and
this is an indication that the dS theory is more relevant than
the Poincare and AdS ones. We believe that for understanding
which of those possibilities are ``better" (if any), other
approaches to quantum theory should be investigated. For
example, in a quantum theory over a Galois field \cite{lev2,lev2B,lev2C},
Poincare symmetry is not possible and even in the AdS case one
IR describes a particle and its antiparticle simultaneously. In
Galois fields the notions of ``less than", ``greater than" and
``positive and negative numbers" can be only approximate. In
particular, there can be no IRs over a Galois field with the
lowest weight or highest weight. As a consequence, in this
theory, as well as in standard dS theory, such quantum numbers
as the electric, baryon and lepton charges can be conserved
only in some approximations and there are no neutral elementary
particles. However, we did not succeed in proving that only
fermions can be elementary since in Galois fields not only
Equation (\ref{54}) is possible but $\eta(j)\eta(j)^*=-1$ is
possible too.

A possible approach for seeking new theories might be based on
finding new symmetries such that known symmetries are special
cases of the new ones when a contraction parameter goes to zero
or infinity (see e.g., the famous paper \cite{Dyson} entitled
``Missed Opportunities"). For example, classical theory is a
special case of quantum one when $\hbar\to 0$ and
non-relativistic theory is a special case of relativistic one
when $c\to\infty$. From this point of view, dS and AdS
symmetries are ``better" than Poincare one since the latter is
a special case of the former when $R\to\infty$. A question
arises whether there exists a ten-dimensional algebra, which is
more general than the dS or AdS one, {\it i.e.}, the dS or AdS
algebra is a special case of this hypothetical new algebra when
some parameter goes to zero or infinity. As noted in Reference
\cite{Dyson}, the answer is ``no" since the dS and AdS algebras
are semisimple. So one might think that the only way to extend
the de Sitter symmetries is to consider higher dimensions and
this is in the spirit of modern trend.

However, if we consider a quantum theory not over complex
numbers but over a Galois field of characteristic $p$ then
standard dS and AdS symmetries can be extended as follows. We
require that the operators $M^{ab}$ satisfy the same
commutation relations but those operators are considered in
spaces over a Galois field. Such operators implicitly depend on
$p$ but they still do not depend on $R$. This approach, which
we call quantum theory over a Galois field (GFQT), has been
discussed in details in References \cite{lev2,lev2B,lev2C,lev3,massless}.
GFQT is a more general theory than standard one since the
latter is a special case of the former when $p\to\infty$. In
the approximation when $p$ is very large, GFQT can reproduce
all the standard results of quantum theory. At the same time,
GFQT is well defined mathematically since it does not contain
infinities. While in standard theory the dS and AdS algebras
are ``better" than the Poincare algebra from aesthetic
considerations (see the discussion in Section \ref{symmetry}}),
in GFQT there is no choice since the Poincare algebra over a
Galois field is unphysical (see the discussion in References
\cite{lev2,lev2B,lev2C,lev3}).

In view of the above discussion, it seems natural to express
all dimensionful quantities in terms of $(c,\hbar,R)$ rather
than $(c,\hbar,G)$ since the former is a set of parameters
characterizing transitions from higher symmetries to lower
ones. Then a reasonable question is why the quantity $G$ is so
small. Indeed, in units $\hbar=c=1$, $G$ has the dimension
$length^2$ and so one might expect that it should be of order
$R^2=3/\Lambda$. So again the disagreement is more that 120
orders of magnitude and one might call this the gravitational
constant problem rather than the cosmological constant problem.
As noted above, in standard theory a reasonable possibility is
that $G\Lambda$ is of order unity. However, in GFQT we have a
parameter $p$. In Reference \cite{essay} we have described our
hypothesis that $G$ contains a factor $1/lnp$ and that is why
it is so small.

In the present paper we have shown that the well known
classical result about the cosmological repulsion in the dS
space is a special case of quantum theory with the dS algebra
as the symmetry algebra when no interaction between particles
is introduced and quasiclassical approximation is valid. Our
result has been obtained without using the notions of spacetime
background, Riemannian geometry and dS QFT. This result shows
that for explaining the fact that $\Lambda>0$ there is no need
to involve dark energy or other fields.

\begin{sloppypar}
The main achievements of modern theory have been obtained in
the approach proceeding from spacetime background. In quantum
theory this approach is not based on a solid mathematical basis
and, as a consequence, the problem of infinities arises. While
in QED and other renormalizable theories this problem can be
somehow circumvented, in quantum gravity this is not possible
even in lowest orders of perturbation theory. Mathematical
problems of quantum theory are discussed in a wide literature.
For example, in the well known textbook \cite{Bogolubov} it is
explained in details that interacting quantized fields can be
treated only as operatorial distributions and hence their
product at the same point is not well defined. One of ideas of
the string theory is that if a point (a zero-dimensional
object) is replaced by a string (a one-dimensional object) then
there is hope that infinities will be less singular.
\end{sloppypar}

For the majority of physicists the fact that GR and quantum
theory describe many experimental data with an unprecedented
accuracy is much more important than a lack of mathematical
rigor, existence of infinities and that the notion of spacetime
background is not physical. For this reason physicists do not
wish to abandon this notion. As one of the consequences, the
cosmological constant problem arises and it is now believed
that dark energy accounts for more than 70\% of the total
energy of the Universe. There exists a vast literature where
different authors propose different approaches and some of the
authors claim that they have found the solution of the problem.
Meanwhile the above discussion clearly demonstrates that the
cosmological constant problem (which is often called the dark
energy problem) is a purely artificial problem arising as a
result of using the notion of spacetime background while this
notion is not physical.

The conclusion that the cosmological constant problem does not
exist has been made earlier by different authors from different
considerations but probably all those authors accepted
approaches based on spacetime background. For example, Bianchi
and Rovelli in their paper \cite{Bianchi} entitled ``Why all
these prejudices against a constant?''  discussed this problem
in the framework of classical GR. They argue that since the
most general form of Einstein equations contains both, $G$ and
$\Lambda$, there is no reason to believe that nature prefers a
special case $\Lambda=0$. In their approach, both $G$ and
$\Lambda$ are fundamental physical quantities. In our paper we
argue that none of this quantity is fundamental and this is in
agreement with recent intensive investigation of a possibility
that gravity in not fundamental but emergent.

\begin{center} {\bf Acknowledgements}\end{center}
\begin{sloppypar}

L.A. Kondratyuk and S.N. Sokolov paid my attention to Dirac's
paper \cite{Dir} and numerous discussions with them have
considerably enhanced my understanding of quantum theory. I
have also greatly benefited from the book \cite{Mensky} by M.B.
Mensky on the theory of induced representations for physicists
\linebreak (I am surprised that this book has been published
only in Russia). E.G. Mirmovich proposed an idea that only
angular momenta are fundamental physical quantities
\cite{Mirmovich}. I am also grateful to Volodya Netchitailo and
Teodor Shtilkind for numerous discussions and to Jim Bogan for
an interesting correspondence and explaining me the
experimental status of the cosmological constant problem.
\end{sloppypar}

\end{document}